\documentclass[twocolumn]{aastex63}
\usepackage{natbib}
\usepackage{tipa}
\usepackage[normalem]{ulem}
\usepackage{calc}
\newcommand\hcancel[2][0.5pt]{%
  \ifmmode\sbox\CBox{$#2$}\else\sbox\CBox{#2}\fi%
  \makebox[0pt][l]{\usebox\CBox}%
  \rule[0.5\ht\CBox-#1/2]{\wd\CBox}{#1}}

\newcommand\kepler{\textit{Kepler}}

\newcommand\gaia{\textit{Gaia}}

\newcommand{\unit}[1]{\ensuremath{\, \mathrm{#1}}} 

\shortauthors{Beard et al. 2022}
\shorttitle{TOI-1696 and TOI-2136: Constraining the Masses of Two Mini-Neptunes with HPF}

\begin{document}

\title{TOI-1696 and TOI-2136: Constraining the Masses of Two Mini-Neptunes with HPF}

\author[0000-0001-7708-2364]{Corey Beard}
\affiliation{Department of Physics and Astronomy, 4129 Frederick Reines Hall, University of California, Irvine, Irvine, CA, 92697, USA}


\author[0000-0003-0149-9678]{Paul Robertson}
\affiliation{Department of Physics and Astronomy, 4129 Frederick Reines Hall, University of California, Irvine, Irvine, CA, 92697, USA}

\author[0000-0001-8401-4300]{Shubham Kanodia}
\affiliation{Department of Astronomy \& Astrophysics, 525 Davey Laboratory, The Pennsylvania State University, University Park, PA, 16802, USA}
\affiliation{Center for Exoplanets and Habitable Worlds, 525 Davey Laboratory, The Pennsylvania State University, University Park, PA, 16802, USA}

\author[0000-0002-2990-7613]{Jessica Libby-Roberts}
\affil{Department of Astronomy \& Astrophysics, 525 Davey Laboratory, The Pennsylvania State University, University Park, PA, 16802, USA}
\affil{Center for Exoplanets and Habitable Worlds, 525 Davey Laboratory, The Pennsylvania State University, University Park, PA, 16802, USA}


\author[0000-0003-4835-0619]{Caleb I. Ca\~nas}
\affiliation{NASA Earth and Space Science Fellow}
\affil{Department of Astronomy \& Astrophysics, 525 Davey Laboratory, The Pennsylvania State University, University Park, PA, 16802, USA}
\affil{Center for Exoplanets and Habitable Worlds, 525 Davey Laboratory, The Pennsylvania State University, University Park, PA, 16802, USA}

\author[0000-0002-5463-9980]{Arvind F.\ Gupta}
\affil{Department of Astronomy \& Astrophysics, 525 Davey Laboratory, The Pennsylvania State University, University Park, PA, 16802, USA}
\affil{Center for Exoplanets and Habitable Worlds, 525 Davey Laboratory, The Pennsylvania State University, University Park, PA, 16802, USA}

\author[0000-0002-5034-9476]{Rae Holcomb}
\affiliation{Department of Physics and Astronomy, 4129 Frederick Reines Hall, University of California, Irvine, Irvine, CA, 92697, USA}

\author[0000-0002-7227-2334]{Sinclaire Jones}
\affiliation{Princeton University, Department of Astrophysical Sciences, 4 Ivy Lane, Princeton, NJ 08540, USA}

\author[0000-0002-4475-4176]{Henry A. Kobulnicky}
\affil{Department of Physics \& Astronomy, University of Wyoming, Laramie, WY 82070, USA}

\author[0000-0002-9082-6337]{Andrea S.J.\ Lin}
\affil{Department of Astronomy \& Astrophysics, 525 Davey Laboratory, The Pennsylvania State University, University Park, PA, 16802, USA}
\affil{Center for Exoplanets and Habitable Worlds, 525 Davey Laboratory, The Pennsylvania State University, University Park, PA, 16802, USA}

\author[0000-0001-8342-7736]{Jack Lubin}
\affiliation{Department of Physics and Astronomy, 4129 Frederick Reines Hall, University of California, Irvine, Irvine, CA, 92697, USA}

\author[0000-0001-8222-9586]{Marissa Maney}
\affiliation{Harvard-Smithsonian Center for Astrophysics, 60 Garden Street, Cambridge, MA 02138, US}

\author[0000-0001-9307-8170]{Brock A. Parker}
\affil{Department of Physics \& Astronomy, University of Wyoming, Laramie, WY 82070, USA}

\author[0000-0001-7409-5688]{Gu\textipa{\dh}mundur Stefánsson}
\newcommand{\Princeton}{Department of Astrophysical Sciences, Princeton University, 4 Ivy Lane, Princeton, NJ 08540, USA}
\newcommand{\RUSSELL}{Henry Norris Russell Fellow}  
\affil{\Princeton}
\affil{\RUSSELL}


\author[0000-0001-9662-3496]{William D. Cochran}
\affiliation{McDonald Observatory and Center for Planetary Systems Habitability, The University of Texas, Austin TX 78712 USA.}

\author[0000-0002-7714-6310]{Michael Endl}
\affiliation{Center for Planetary Systems Habitability, The University of Texas at Austin, Austin, TX 78712, USA}
\affiliation{Department of Astronomy, The University of Texas at Austin, TX, 78712, USA}

\author[0000-0003-1263-8637]{Leslie Hebb}
\affiliation{Department of Physics, Hobart and William Smith Colleges, 300 Pulteney Street, Geneva,
NY, 14456, USA}

\author[0000-0001-9596-7983]{Suvrath Mahadevan}
\affil{Department of Astronomy \& Astrophysics, 525 Davey Laboratory, The Pennsylvania State University, University Park, PA, 16802, USA}
\affil{Center for Exoplanets and Habitable Worlds, 525 Davey Laboratory, The Pennsylvania State University, University Park, PA, 16802, USA}

\author[0000-0001-9209-1808]{John Wisniewski}
\affiliation{Homer L. Dodge Department of Physics and Astronomy, University of Oklahoma, 440 W. Brooks Street, Norman, OK 73019, USA}


\author[0000-0003-4384-7220]{Chad F. Bender}
\affil{Steward Observatory, The University of Arizona, 933 N.\ Cherry Avenue, Tucson, AZ 85721, USA}

\author[0000-0002-2144-0764]{Scott A. Diddams}
\affiliation{National Institute of Standards \& Technology, 325 Broadway, Boulder, CO 80305, USA}
\affiliation{Department of Physics, 390 UCB, University of Colorado Boulder, Boulder, CO 80309, USA}

\author[0000-0002-0885-7215]{Mark Everett}
\affiliation{National Optical Infrared Astronomy Research Laboratory 950 N. Cherry Ave., Tucson, AZ 85719}

\author[0000-0002-0560-1433]{Connor Fredrick}
\affil{Time and Frequency Division, National Institute of Standards and Technology, 325 Broadway, Boulder, CO 80305, USA}
\affil{Department of Physics, University of Colorado, 2000 Colorado Avenue, Boulder, CO 80309, USA}

\author[0000-0003-1312-9391]{Samuel Halverson}
\affil{Jet Propulsion Laboratory, 4800 Oak Grove Drive, Pasadena, CA 91109, USA}

\author[0000-0002-1664-3102]{Fred Hearty}
\affil{Department of Astronomy \& Astrophysics, 525 Davey Laboratory, The Pennsylvania State University, University Park, PA, 16802, USA}
\affil{Center for Exoplanets and Habitable Worlds, 525 Davey Laboratory, The Pennsylvania State University, University Park, PA, 16802, USA}

\author[0000-0001-5000-1018]{Andrew J. Metcalf}
\affiliation{Space Vehicles Directorate, Air Force Research Laboratory, 3550 Aberdeen Ave. SE, Kirtland AFB, NM 87117, USA}
\affiliation{Time and Frequency Division, National Institute of Technology, 325 Broadway, Boulder, CO 80305, USA} 
\affiliation{Department of Physics, 390 UCB, University of Colorado Boulder, Boulder, CO 80309, USA}

\author[0000-0002-0048-2586 ]{Andrew Monson}
\affiliation{Department of Astronomy \& Astrophysics,  525 Davey Laboratory, The Pennsylvania State University,  University Park, PA, 16802, USA}

\author[0000-0001-8720-5612]{Joe P.\ Ninan}
\affil{Department of Astronomy \& Astrophysics,  525 Davey Laboratory, The Pennsylvania State University,  University Park, PA, 16802, USA}
\affil{Center for Exoplanets and Habitable Worlds,  525 Davey Laboratory, The Pennsylvania State University, University Park, PA, 16802, USA}

\author[0000-0001-8127-5775]{Arpita Roy}
\affiliation{ Space Telescope Science Institute, 3700 San Martin Drive, Baltimore, MD 21218, USA}
\affiliation{Department of Physics and Astronomy, Johns Hopkins University, 3400 N Charles St, Baltimore, MD 21218, USA}

\author[0000-0003-2435-130X]{Maria Schutte}
\affiliation{Homer L. Dodge Department of Physics and Astronomy, University of Oklahoma, 440 W. Brooks Street, Norman, OK 73019, USA}

\author[0000-0002-4046-987X]{Christian Schwab}
\affil{Department of Physics and Astronomy, Macquarie University, Balaclava Road, North Ryde, NSW 2109, Australia}

\author[0000-0002-4788-8858]{Ryan C Terrien}
\affiliation{Carleton College, One North College St., Northfield, MN 55057, USA}

\correspondingauthor{Corey Beard}
\email{coreycbeard@gmail.com}

\begin{abstract}

 We present the validation of two planets orbiting M dwarfs, TOI-1696b and TOI-2136b. Both planets are mini-Neptunes orbiting nearby stars, making them promising prospects for atmospheric characterization with the James Webb Space Telescope. We validated the planetary nature of both candidates using high contrast imaging, ground-based photometry, and near-infrared radial velocities. Adaptive Optics images were taken using the ShARCS camera on the 3 m Shane Telescope. Speckle images were taken using the NN-Explore Exoplanet Stellar Speckle Imager on the WIYN 3.5 m telescope. Radii and orbital ephemerides were refined using a combination of TESS, the diffuser-assisted ARCTIC imager on the 3.5m ARC telescope at Apache Point Observatory, and the 0.6 m telescope at Red Buttes Observatory. We obtained radial velocities using the Habitable-Zone Planet Finder on the 10 m Hobby-Eberly Telescope, which enabled us to place upper limits on the masses of both transiting planets. TOI-1696b (P = 2.5 days; R$_{p}$ = 3.24 R$_{\oplus}$; M$_{p}$ $<$ 56.6 M$_{\oplus}$) falls into a sparsely-populated region of parameter space considering its host star's temperature (T$_{\rm{eff}}$ = 3168 K, M4.5), as planets of its size are quite rare around mid to late M dwarfs. On the other hand, TOI-2136b (P = 7.85 days; R$_{p}$ = 2.09 R$_{\oplus}$; M$_{p}$ $<$ 15.0 M$_{\oplus}$) is an excellent candidate for atmospheric follow-up with JWST.

\end{abstract}


\section{Introduction} \label{sec:intro}

Exoplanet Discoveries were once rare and challenging, but today our methods have advanced sufficiently to attempt not just discovering exoplanets, but understanding them. The \kepler\ mission \citep{borucki10} revolutionized the field, providing researchers with a statistical sample of exoplanets. We stand on the verge of answering some of the biggest questions in exoplanet astronomy: how do planets form? What is the true population of exoplanets? Is there life on any of these planets? With the successful launch of the \textit{James Webb Space Telescope} (JWST; \citealt{gardner06}), and the advent of other ultra-precise instruments under development, the answers to these questions are near at hand.

Today, the \textit{Transiting Exoplanet Survey Satellite} (TESS) is providing near-constant high precision photometry of thousands of bright, nearby stars \citep{ricker2015}. TESS has already flagged more than 5000 TESS Objects of Interest (TOIs) as potential exoplanets, but additional follow-up is necessary. The transit method is especially susceptible to false positive scenarios \citep{fressin13}, and additional data is required to verify the planetary nature of TOIs. This is even more important for TESS observations: a pixel size of 21\unit{\arcsec} is so large that TESS pixels almost always contain multiple stars. Beyond ruling out false positives, additional ground-based follow-up is necessary for determining other planetary properties, especially if we wish to understand the astrophysics behind formation, evolution, and populations of exoplanets.

The radial velocity (RV) method allows us to measure masses of exoplanets, and thus, in combination with derived radii from photometry, constrain properties such as bulk density and composition. Previously, RV instruments (Keck/HIRES \citep{vogt1994}, HARPS \citep{mayor03}, HARPS-N \citep{cosentino2012}) focused on the visible wavelength range, often designed primarily to study Sun-like stars. The existence of exoplanets around stars of all spectral types, however, has broadened our search, in particular to the redder, cooler M dwarf spectral type: the most common type of star in our galaxy \citep{henry18}. The extended main-sequence lifetime of these stars is thought contribute to the probability of life forming \citep{shields16}, making biosignature searches potentially more fruitful. Thus we see the development of dedicated infrared RV instruments, i.e. the Habitable-Zone Planet Finder (HPF) \citep{mahadevan12, mahadevan14}, iSHELL \citep{cale18}, IRD \citep{tamura12} PARVI \citep{gibson20}, SPIRou \citep{donati18}, and NIRPS \citep{bouchy17}; instruments with visible and infrared observing capabilities, i.e.  CARMENES \citep{quirrenbach16} or GIARPS \citep{claudi17}; or instruments that are either redder than traditional visible spectrographs, or have wider coverage i.e. NEID \citep{schwab16} or ESPRESSO \citep{pepe10}. 

Evidence that the M dwarf exoplanet population is distinct from that of Sun-like stars motivates further study of this group \citep{dressing15}. In particular, the mass-radius relationship for M dwarfs is less well understood, and additional mass measurements will help us to explain its underlying form \citep{kanodia19}.

Here we validate two transiting M dwarf planets identified by TESS: TOI-1696b \citep{mori21} and TOI-2136b. We also use HPF to place upper limits on the masses of both planets. We additionally refine the measured planetary, orbital, and stellar parameters of both systems.

TOI-1696 is a mid M dwarf (M4.5, V=16.8) with particularly deep transit events. The cool, small nature of the star and the relatively large transit depth suggest that this planet is an unusually large ($\sim$ 3 R$_{\oplus}$) mini-Neptune considering its host star's temperature (T$_{\rm{eff}}$=3168 K)---planets larger than 2.8 R$_{\oplus}$ become quite rare around cool stars \citep{dressing15, hardegree-ullman19, hsu20}. \cite{mori21} highlighted several attractive features of the system: it has a high Transmission Spectroscopy Metric (TSM; \citealt{kempton18}) considering the cool nature of the star, and the planet is near the Neptunian desert \citep{mazeh16}. An exoplanet that is unusually large, approaching the Neptunian desert, and that can be atmospherically studied can help explain or rule out planet formation scenarios, and our understanding of planetary evolution.

TOI-2136 is an early to mid M dwarf (M3, T$_{\rm{eff}}$ = 3366 K, V=14.3). The candidate planet is a small mini-Neptune ($\sim$ 2 R$_{\oplus}$); with an estimated equilibrium temperature of 403 K, TOI-2136b falls into a regime that allows for the existence of liquid water oceans beneath its gaseous envelope. TOI-2136's bright, close nature also puts this candidate planet in a potentially exciting region of parameter space for the detection of biosignatures: For example, TOI-2136b is an excellent candidate for a ``cold Haber World," a unique environment where life could exist in oceans beneath a massive H/He envelope by combining hydrogen and nitrogen in the ``Haber" process \citep{seager13a}. Such a process generates a detectable amount of ammonia, and TOI-2136b is not expected to produce ammonia in any other way.

In Section \ref{sec:observations}, we give a summary of the data used in our analysis. In Section \ref{sec:stellar}, we detail our estimation of each system's stellar parameters. In Section \ref{sec:analysis}, we detail the software and assumptions made during our analysis, and steps taken to measure the planetary and orbital parameters. In Section \ref{sec:discussion}, we discuss our findings, and the implications for each system. Finally, Section \ref{sec:summary} summarizes our results and conclusions.

\section{Observations}\label{sec:observations}
\subsection{TESS}\label{sec:TESS}

\begin{figure*}[] 
\centering
\includegraphics[width=\textwidth]{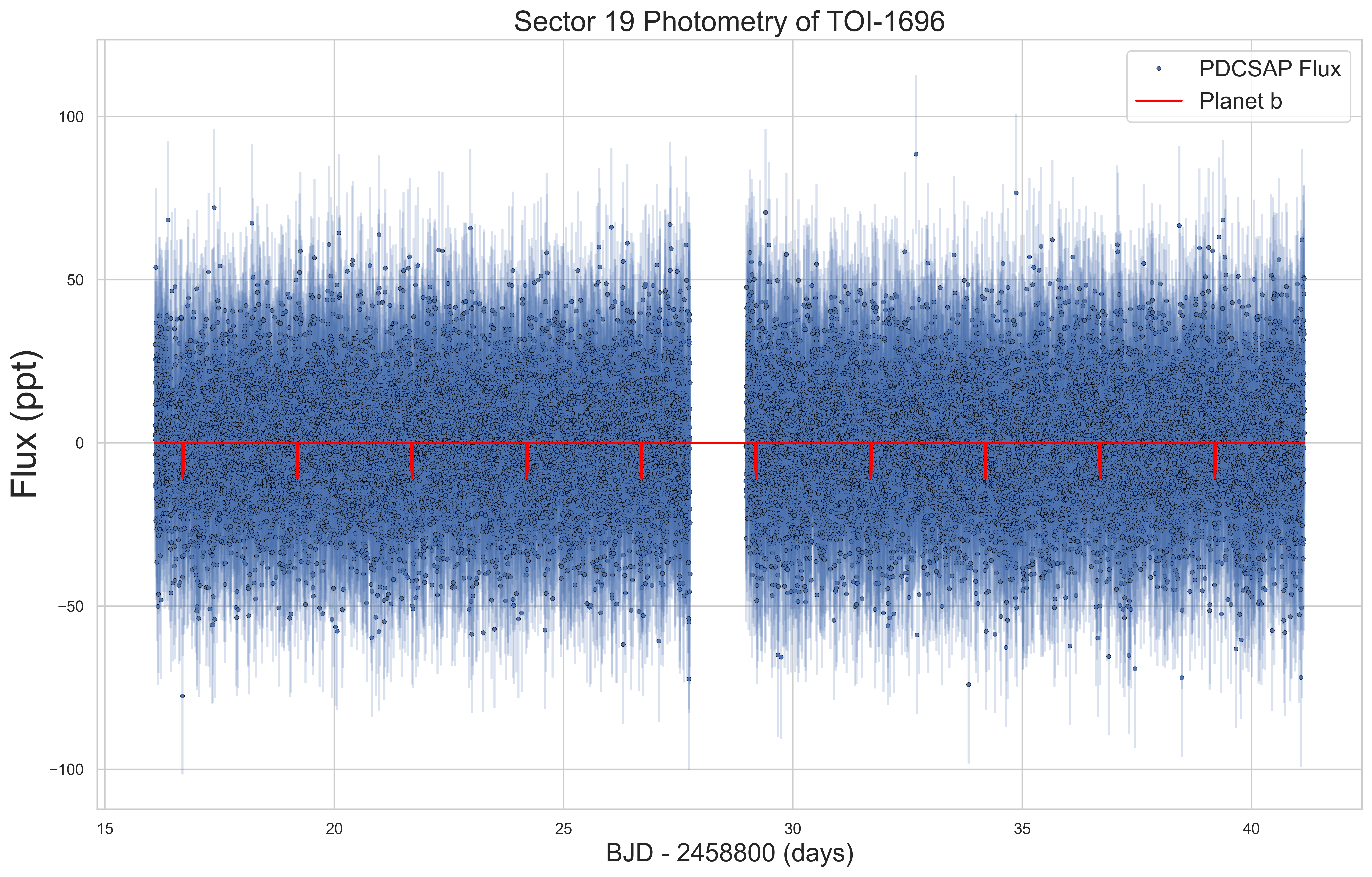}
\caption{Sector 19 TESS PDCSAP flux of TOI-1696, reduced using the SPOC pipeline. A model representing the transits of planet b is indicated by a solid red line.} \label{fig:sector19photometry}
\end{figure*}

TOI-1696 was observed by TESS during Sector 19 (2019 November 27 - 2019 December 24)\footnote{1696 DV: https://tev.mit.edu/data/atlas-signal/i165570/}. It was observed for 27 days by CCD 2 of camera 1 in 2-minute cadence mode. It was then processed by the MIT Science Processing Operations Center (SPOC) pipeline \citep{jenkins16}, and was announced as a TOI on January 25, 2020, identifying a 2.5 day periodicity as TOI-1696.01. Photometry taken during Sector 19 is shown in Figure \ref{fig:sector19photometry}.

\begin{figure*}[] 
\centering
\includegraphics[width=\textwidth]{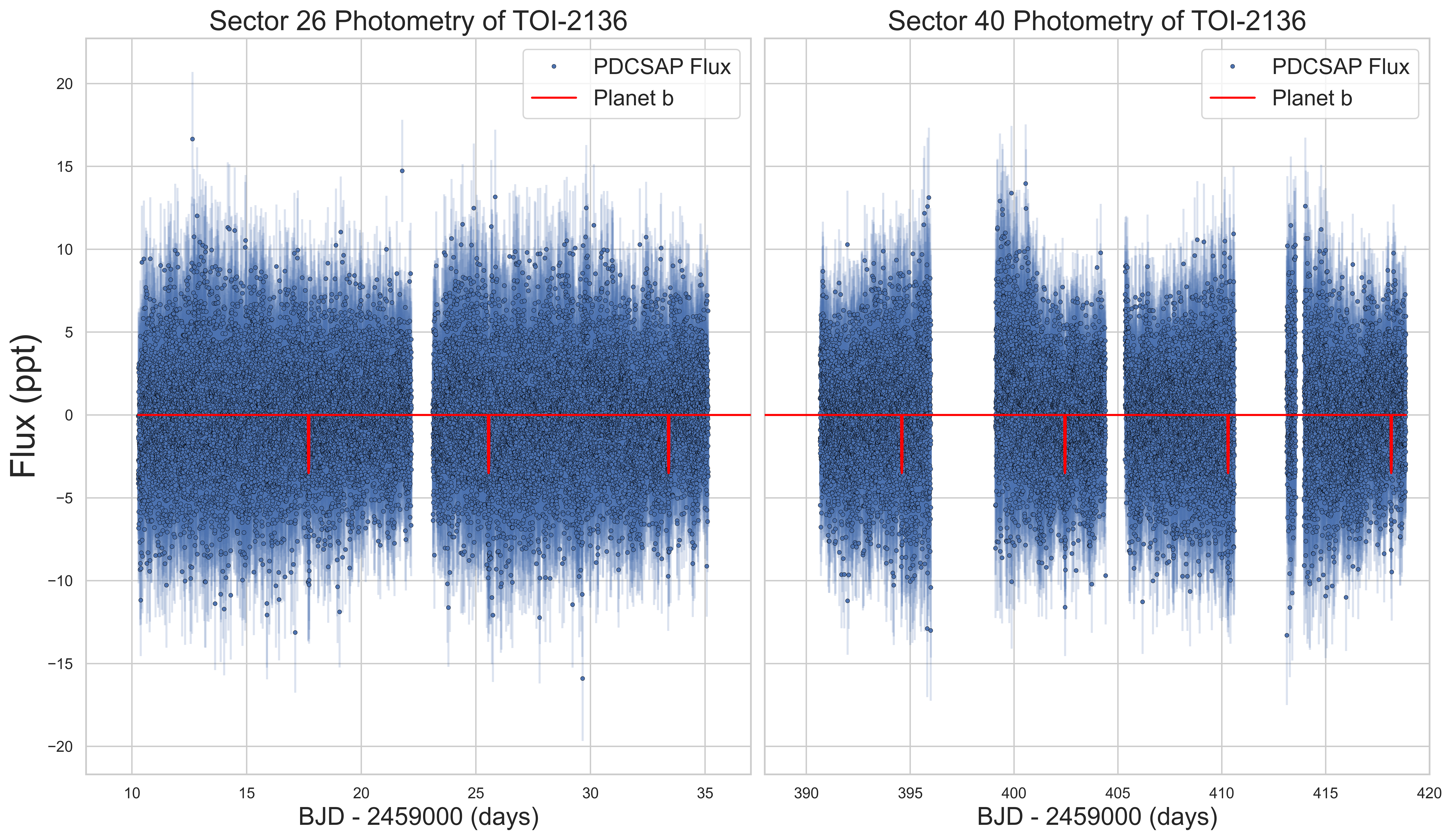}
\caption{Sector 26 and 40 PDCSAP flux of TOI-2136. A model representing the transits of planet b is visible as a solid red line.} \label{fig:sector26_40photometry}
\end{figure*}

Similarly, TOI-2136 was observed in Sector 26 (2020 June 8th - 2020 July 4th) during TESS's nominal mission, and Sector 40 (2021 June 24th - 2021 July 23rd) during TESS's extended mission, in 2-minute cadence mode with CCD 1 of camera 1\footnote{2136 DV: https://tev.mit.edu/data/atlas-signal/i177995/}. A 7.85 day planet candidate was identified on 2020 August 27 as TOI-2136.01. The target is also listed in Sector 14 by the Web-TESS Viewing Tool, but follow-up using the \textsf{TESS-point} \citep{Burke20} software package confirms that the target fell in a gap between CCDs during that Sector. Both observed sectors were processed using the SPOC pipeline and used for our subsequent analysis. We use the Pre-Search Data Conditioning Simple Aperture Photometry (PDCSAP) flux in our analysis. A plot of both sectors of TESS photometry for TOI-2136 is visible in Figure \ref{fig:sector26_40photometry}.

It should be noted that the Sector 40 photometry of TOI-2136 has several large gaps. The central data gap at $\sim$ BJD 2459405 is a standard data downlink. The other gaps, most notably BJD 2459396-2459399, BJD 2459411-2459413, and a small gap around BJD 2459414 are all due to an attitude adjustment of the spacecraft as described in the TESS Mission Handbook. These data were not used during the analysis.

\subsection{Ground Based Photometric Follow-up}\label{sec:gbphotometry}

Ground based photometric follow-up is often necessary to validate the planetary nature of candidates flagged by TESS as TOIs \citep{stefansson20_b, canas22}. In addition to confirming signals, these follow-up transits provide tighter constraints on transit parameters for the candidate planets (e.g. \cite{kanodia21}). We detail the photometric follow-up for these two planet candidates using ground based resources in the next section.

\subsubsection{RBO}\label{sec:rbo}

We observed a transit of TOI-1696 on the night of 2020 December 27 using the $0.6~\unit{m}$ telescope at Red Buttes Observatory (RBO) in Wyoming \citep{kasper_remote_2016}. The RBO telescope is a f/8.43 Ritchey-Chrétien Cassegrain constructed by DFM Engineering, Inc.  It is currently equipped with an Apogee ASPEN CG47 camera.

The target rose from an airmass of 1.092 at the start of the observations to a minimum airmass of 1.009, and then set to an airmass of 1.419 at the end of the observations. Observations were performed using the Bessell I filter with $2\times2$ on-chip binning. We defocused moderately, which allowed us to use an exposure time of 120$ \unit{s}$. In the $2\times2$ binning mode, the 0.6$ \unit{m}$ at RBO has a gain of 1.27$ \unit{e/ADU}$, a plate scale of 1.05$ \arcsec$, and a readout time of approximately 2$ \unit{s}$.

During data reduction we used a 9 pixel (9.5\unit{\arcsec}) aperture and an annulus with an inner radius of 16 pixels (16.8\unit{\arcsec}) and an outer radius of 24 pixels (25.2\unit{\arcsec}).

We note that the latter observations seem to be affected by particularly high scatter. Analysis of airmass correlation doesn't resolve this issue, and so we conclude that the observations were possibly affected by clouds or poor seeing towards the end. The high scatter does not have a significant effect on our measured parameters, however, as it seems to be concentrated after egress. We leave the points for completeness. 

Plots of RBO data are visible in Section \ref{sec:transitanalysis}, where they are analyzed.

\subsubsection{ARCTIC}\label{sec:arctic}

We observed  transits of TOI-1696 and TOI-2136 on the nights of 2021 January 1 and 2020 October 5, respectively, using the Astrophysical Research Consortium (ARC) Telescope Imaging Camera \citep[ARCTIC;][]{huehnerhoff_astrophysical_2016} at the ARC 3.5 m Telescope at Apache Point Observatory (APO). To achieve precise photometry on nearby bright stars, we used the engineered diffuser described in \citet{stefansson_toward_2017}.

The airmass of TOI-1696 varied from 1.040 to 1.384 over the course of its observation on 2021 January 1. The observations were performed using the SDSS $i^{\prime}$ filter with an exposure time of  20$\unit{s}$ in the quad-readout and fast readout modes with $4\times4$ on-chip binning. In the $4\times4$ binning mode, ARCTIC has a gain of 2\unit{e/ADU}, a plate scale of 0.468 \unit{\arcsec/pixel}, and a readout time of 2.7 $\unit{s}$. We initially defocused to a FWHM of 4.4\unit{\arcsec}. For the final reduction, we selected a photometric aperture of 8 pixels (3.7\unit{\arcsec}) and used an annulus with an inner radius of 20 pixels (9.4\unit{\arcsec}), and an outer radius of 30 pixels (14.0\unit{\arcsec}).

The airmass of TOI-2136 varied from 1.016 to 1.420 over the course of its observation on 2020 October 5. The observations were performed using the SDSS i$^{\prime}$ filter with an exposure time of 15.3$\unit{s}$ in the quad-readout and fast readout modes with $4\times4$ on-chip binning. For the final reduction, we selected a photometric aperture of 17 pixels (8.0\unit{\arcsec}), and an annulus with an inner radius of 32 pixels (15.0\unit{\arcsec}) and an outer radius of 48 pixels (22.5\unit{\arcsec}).

Plots of ARCTIC data are visible in Section \ref{sec:transitanalysis}, where they are analyzed.

\subsection{Radial Velocity Follow-Up with the Habitable Zone Planet Finder}\label{sec:hpfrvs}

We observed both targets using HPF \citep{mahadevan12, mahadevan14}, a near-infrared (\(8080-12780\)\ \AA), high precision RV spectrograph located at the 10 m Hobby-Eberly Telescope (HET) in Texas. HET is a fixed-altitude telescope with a roving pupil design. Observations on the HPF are queue-scheduled, with all observations executed by the HET resident astronomers \citep{shetrone07}. HPF is fiber-fed, with separate science, sky and simultaneous calibration fibers \citep{kanodia18a}, and has precise, milli-Kelvin-level thermal stability \citep{stefansson16}. 

To estimate the RVs, we use a modified version of the \texttt{SpEctrum Radial Velocity AnaLyser} pipeline \citep[\texttt{SERVAL};][]{zechmeister18}, reduced using the method outlined in \cite{stefansson20_a}. \texttt{SERVAL} matches templates to the obtained spectra to create a master template from all observations, and then minimizes the \(\chi^2\) statistic to determine the shifts of each observed spectrum. This method is widely used for M dwarfs, where line blends make the binary mask technique less effective \citep[e.g.,][]{anglada-escude12}. We create a  master template for each target from all it observed spectra. We mask telluric and sky-emission lines during this process. We calculate a telluric mask based on their predicted locations using \texttt{telfit} \citep{gullikson14}, a Python wrapper to the Line-by-Line Radiative Transfer Model package \citep{clough05}.  Despite their proximity to Earth, both targets are relatively faint. As a result, sky-fiber spectra were subtracted from the observations. Analyses were run on both sky-subtracted and non-sky-subtracted RVs to ascertain the effect of this correction. The final analysis results did not differ meaningfully between the runs. We adopt sky-subtracted RVs for both systems due to their faintness, and the long exposure times of both targets. We use \texttt{barycorrpy} to perform barycentric corrections \citep{kanodia18b}.

RVs of TOI-1696 were obtained between 2020 September 27 and 2021 February 25. During this interval we obtained 30 unbinned RVs, taken over 10 observing nights. Each unbinned spectrum was observed for 650 seconds, with an average signal to noise ratio of 33.8. The average RV error was 34.7 m s$^{-1}$ (unbinned) and 19.9 m s$^{-1}$ (binned). The nightly binned RVs are visible in Table \ref{tab:1696rv}.

RVs of TOI-2136 were obtained between 2020 August 13 and 2021 September 19. We obtained 81 unbinned RVs of this system, taken over 27 observing nights. Each observation lasted for 650 seconds. The average signal to noise ratio was 81.7. The mean uncertainty of all of the observations is 14.1 m s$^{-1}$ (unbinned) and 8.0 m s$^{-1}$ (binned). A truncated list of nightly binned RVs can be seen in Table \ref{tab:2136rv}.

\begin{deluxetable}{cccc}
\tablecaption{HPF RVs of TOI-1696. \label{tab:1696rv}}
\tablehead{\colhead{$\unit{BJD_{TDB}}$}  &  \colhead{RV}   & \colhead{$\sigma$} \\
           \colhead{(d)}   &  \colhead{(m s$^{-1}$}) & \colhead{(m s$^{-1}$)} }
\startdata
2459119.84652 & 8 & 16\\
2459124.83479 & -19 & 14\\
2459182.67317 & 34 & 17\\
2459187.88823 & -27 & 19\\
2459210.59442 & 4 & 25\\
2459216.79185 & 15 & 22\\
2459232.76200 & 29 & 17\\
2459237.74584 & 44 & 19\\
2459238.74103 & 23 & 28\\
2459270.64775 & -65 & 21\\
\enddata
\end{deluxetable}

\begin{deluxetable}{cccc}
\tablecaption{HPF RVs of TOI-2136. \label{tab:2136rv}}
\tablehead{\colhead{$\unit{BJD_{TDB}}$}  &  \colhead{RV}   & \colhead{$\sigma$} \\
           \colhead{(d)}   &  \colhead{(m s$^{-1}$}) & \colhead{(m s$^{-1}$)} }
\startdata
2459074.79705 & -13 & 7 \\
2459087.75569 & 17 & 6 \\
2459089.75413 & 9 & 6 \\
2459090.74571 & 0 & 8 \\
2459095.73448 & -1 & 6 \\
2459098.72963 & 5 & 8 \\
2459118.68079 & 16 & 8 \\
2459123.65313 & -1 & 5 \\
2459125.65859 & -8 & 6 \\
2459129.63709 & -11 & 7 \\
\enddata
\end{deluxetable}

\subsection{High Resolution Imaging}

\subsubsection{ShARCS on the Shane Telescope}

We observed TOI-1696 and TOI-2136 using the ShARCS camera on the Shane 3 m telescope at Lick Observatory \citep{srinath14}. TOI-1696 was observed using the $K_S$ filter on the night of 2020 November 29. TOI-2136 was observed using the $K_S$ filter and $J$ filter on the night of 2021 May 28. Both targets were in a brightness regime where Laser Guide Star (LGS) mode is helpful, but this function was unavailable on both nights due to instrument repairs. Fortunately, conditions were good enough in both cases, and Natural Guide Star (NGS) mode proved to be sufficient. Both targets were observed using a 5 point dither process as outlined in \cite{furlan17}.

The raw data are reduced using a custom pipeline developed by our team \citep{stefansson20_b,kanodia21}. Our reduction first rejects all overexposed or underexposed images, and we manually reject files we know to be erroneous from our night logs (lost guiding, shutters in frame, etc). We then apply a standard dark correction, flat correction, and sigma clip. We produce a master sky image from the 5 point dither process, and subtract this sky from each image. A final image is then produced using an interpolation process to shift the images onto a single centroid.

We then generate a 5 sigma contrast curve using algorithms developed by \cite{espinoza16} as the final part of our analysis. For TOI-1696, we detect no companions at a $\Delta$Ks = 2.3 at 0.3$\arcsec$ and $\Delta$Ks = 5.4 at 5.9$\arcsec$. For TOI-2136, we rule out companions with a $\Delta$Ks = 2.4 and a $\Delta$J = 3.3 at 0.2$\arcsec$, and out to $\Delta$Ks = 7.8 and $\Delta$J = 8.7 at a distance of 5.9$\arcsec$.

\begin{figure*}[] 
\centering
\includegraphics[width=\textwidth]{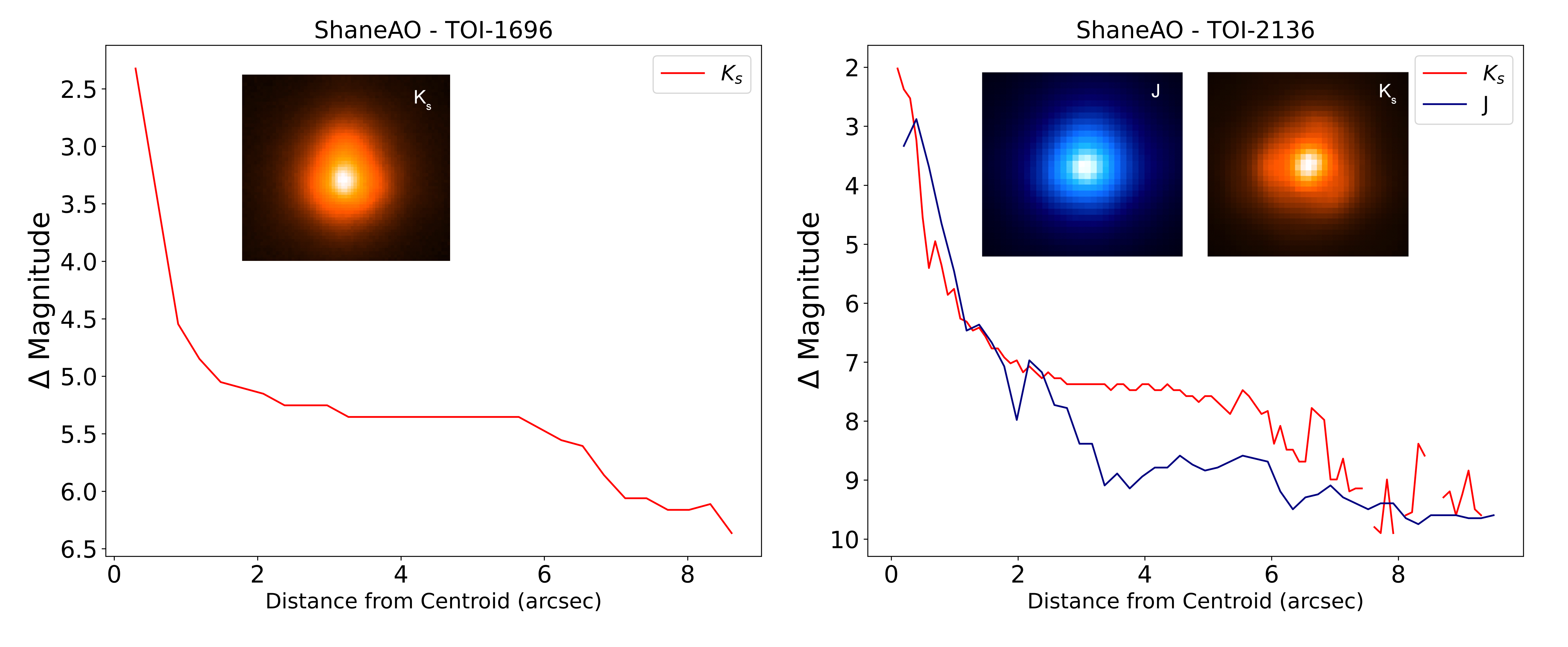}
\caption{Left: 5$\sigma$ contrast curve of TOI-1696 taken using the ShARCS camera at Lick Observatory on 2020 November 29 in the K$_{s}$ band. Right: 5$\sigma$ contrast curves of TOI-2136 taken using the K$_{s}$ and J filters. The data were taken on 2021 May 28.} \label{fig:shane}
\end{figure*}

\subsubsection{NESSI at WIYN}

\begin{figure}[] 
\centering
\includegraphics[width=0.46\textwidth]{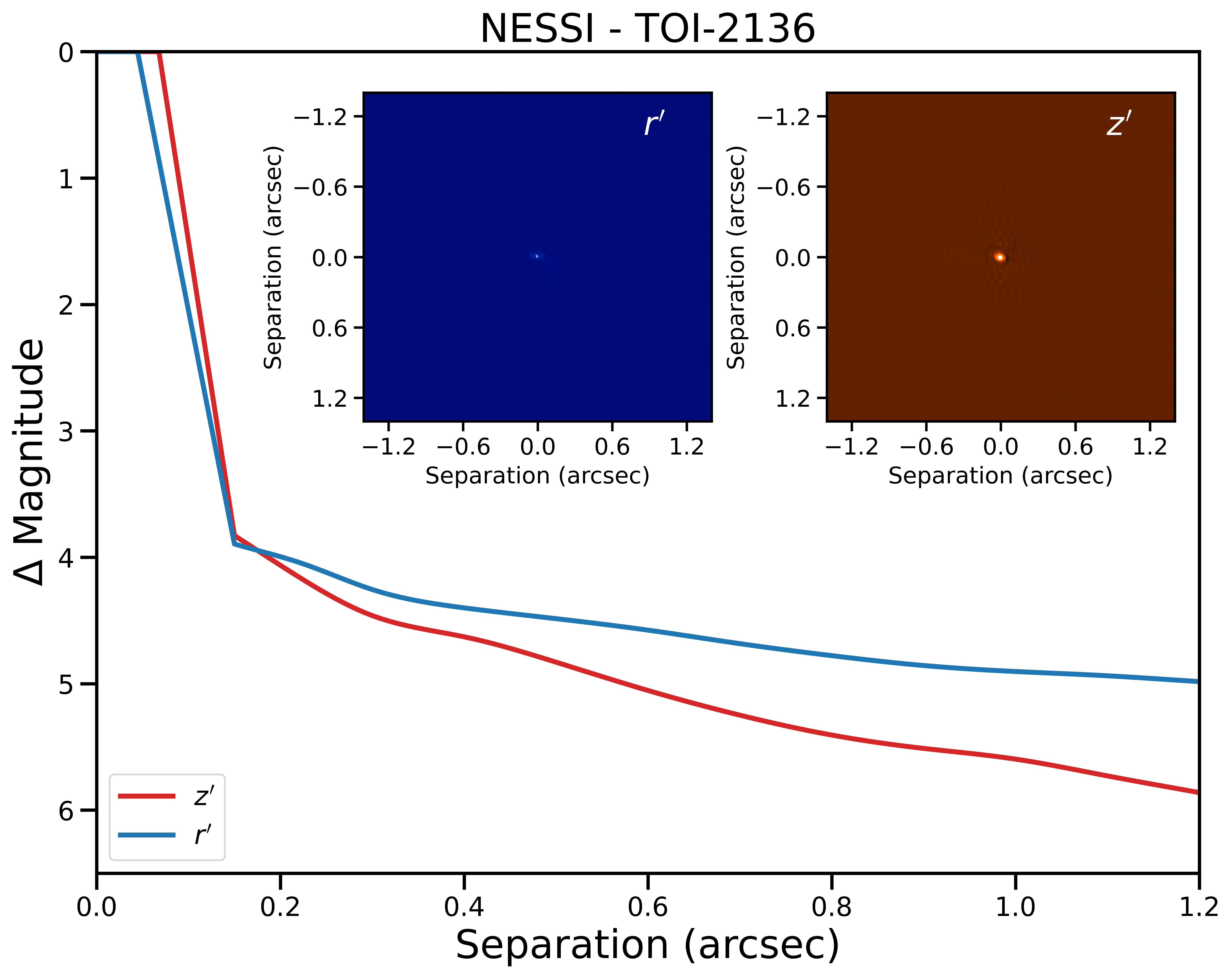}
\caption{Images and contrast curves of TOI-2136 taken using the NN-Explore Exoplanet Stellar Speckle Imager (NESSI) on April 1, 2021. Data were taken in the $r'$ and $z'$ filters, and companions were ruled out to a distance of $1\arcsec$. } \label{fig:nessi2136}
\end{figure}

In addition to Shane AO data, we acquired high-contrast imaging data for TOI-2136 with speckle imaging observations taken on 1 April 2021 using the NN-Explore Exoplanet Stellar Speckle Imager (NESSI) on the WIYN\footnote{The WIYN Observatory is a joint facility of the NSF's National Optical-Infrared Astronomy Research Laboratory, Indiana University, the University of Wisconsin-Madison, Pennsylvania State University, the University of Missouri, the University of California-Irvine, and Purdue University.} 3.5m telescope at Kitt Peak National Observatory. To rule out additional close, luminous companions, we collected a 9 minute sequence of 40 ms diffraction-limited exposures of TOI-2136 with the $r'$ and $z'$ filters. As we show in Figure \ref{fig:nessi2136}, the NESSI data show no evidence of blending from a bright companion down to a contrast limit of $\Delta$mag $=4$ at $0.2\arcsec$ and $\Delta$mag$=5.5$ at $1\arcsec$.

\section{Stellar Parameters}\label{sec:stellar}

We used a similar method to that in \cite{stefansson20_a}, Jones et al. (2022, in prep) to estimate $T_{\mathrm{eff}}$, $\log g$, and [Fe/H] values of the host stars from their spectra. The \texttt{HPF-SpecMatch} code, based on the \texttt{SpecMatch-Emp} algorithm from \cite{yee17}, compares the high resolution HPF spectra of both targets to a library of high SNR as-observed HPF spectra, which consists of slowly-rotating reference stars with well characterized stellar parameters from \cite{yee17} and an expanded selection of stars from \cite{mann15} in the lower effective temperature range. 

We shift the observed target spectrum to a library wavelength scale and rank all of the targets in the library using a $\chi^2$ goodness-of-fit metric. After this initial $\chi^2$ minimization step, we pick the five best matching reference spectra for each target: GJ 3991, PM J08526+2818, GJ 402, GJ 3378, and GJ 1289 in the case of TOI-1696; GJ 251, GJ 581, GJ 109, GJ 436, and GJ 4070 in the case of TOI-2136. From these, we construct a weighted spectrum using their linear combination to better match the target spectrum (Jones et al., 2022, in prep). A weight is assigned to each of the five spectra for each respective target. We then assign the target stellar parameter $T_{\mathrm{eff}}$, $\log g$, and \replaced{Fe/H}{[Fe/H]} values as the weighted average of the five best stars using the best-fit weight coefficients. Our final parameters are listed in Table \ref{stellartable}. These parameters were derived from the HPF order spanning 8670\AA - 8750\AA.

We artificially broadened the library spectra with a $v \sin i$ broadening kernel \citep{gray92} to match the rotational broadening of the target star. We determined both TOI-1696 and TOI-2136 to have $v \sin i$ broadening values of $<$ 2 \unit{km/s}.

We used \texttt{EXOFASTv2} \citep{eastman13} to model the spectral energy distributions (SED) of both systems to derive model-dependent constraints on the stellar mass, radius, and age of each star. \texttt{EXOFASTv2} utilizes the BT-NextGen stellar atmospheric models \citep{allard12} during SED fits.  Gaussian priors were used for the 2MASS (\(JHK\)), SDSS (\(g^\prime r^\prime i^\prime\)), Johnson (\(B V\)), and \textit{Wide-field Infrared Survey Explorer} magnitudes (WISE; $W1$, $W2$, $W3$, and $W4$) \citep[][]{wright10}. Our spectroscopically-derived host star effective temperatures, surface gravities, and metallicities, were used as priors during the SED fits as well, and the estimates from \cite{bailer-jones21} were used as priors for distance. We utilize estimates of Galactic dust by \cite{green19} to estimate the visual extinction for each system. We convert this upper limit to a visual magnitude extinction using the \(R_{v}=3.1\) reddening law from \cite{fitzpatrick99}. Our final model results are visible in Table \ref{stellartable}.

\begin{deluxetable*}{lcccc}
\tablecaption{Stellar Parameters for TOI-1696 and TOI-2136 \label{stellartable}}
\tablehead{\colhead{~~~Parameter Name} &
\colhead{Description} & \colhead{TOI-1696} & \colhead{TOI-2136} & \colhead{Reference$^{a}$}
}
\startdata
\sidehead{Identifiers}
~~~TOI  & TESS Object of Interest & 1696 & 2136 & TESS Mission\\
~~~TIC  & TESS Input Catalog & 470381900 & 336128819 & TICv8 \\
~~~Gaia & GAIA Mission & 270260649602149760 & 2096535783864546944 & Gaia EDR3\\ 
~~~2MASS & 2MASS Identifier & J04210733+4849116 & 18444236+3633445 & 2MASS\\
\sidehead{Coordinates}
~~~$\alpha_{J2016}$  & Right Ascension (deg) & 65.28065076(0) & 281.17633745(6) & Gaia EDR3 \\
~~~$\delta_{J2016}$  & Declination (deg) & 48.81982851(7) & 36.56315642(6) & Gaia EDR3\\
~~~$\mu_{\alpha}$ & Proper Motion RA (mas yr$^{-1}$) & 12.87$\pm$0.03 & -33.80$\pm$0.02 & Gaia EDR3 \\ 
~~~$\mu_{\delta}$ & Proper Motion DEC (mas yr$^{-1}$) & -19.04$\pm$0.03 & 177.05$\pm$0.02 & Gaia EDR3 \\ 
\sidehead{Magnitudes}
~~~V & Johnson V Magnitude & 16.8$\pm$1.1 & 14.3$\pm$0.2 & TICv8 \\
~~~B & Johnson B Magnitude & 18.5$\pm$0.2 & 15.8$\pm$0.1 & TICv8, APASS DR10 \\ 
~~~J & J-band Magnitude & 12.23$\pm$0.02 & 10.18$\pm$0.02 & TICv8 \\
~~~H & H-band Magnitude & 11.60$\pm$0.03 & 9.60$\pm$0.03 & TICv8 \\
~~~K$_{s}$ & K$_{s}$-band Magnitude & 11.33$\pm$0.02 & 9.34$\pm$0.02 & TICv8 \\
~~~g$^{\prime}$ & Sloan g$^{\prime}$ Magnitude & 17.58$\pm$0.03 & 14.9$\pm$0.2 & APASS DR10 \\
~~~r$^{\prime}$ & Sloan r$^{\prime}$ Magnitude & 16.2$\pm$0.1 & 13.2$\pm$0.3 & APASS DR10 \\
~~~i$^{\prime}$ & Sloan i$^{\prime}$ Magnitude & 14.61$\pm$0.03 & 11.9$\pm$ 0.3 & APASS DR10 \\
~~~W1 & WISE 1 Magnitude & 11.13$\pm$0.02 & 9.19$\pm$0.02 & TICv8 \\ 
~~~W2 & WISE 2 Magnitude & 10.98$\pm$0.02 & 9.05$\pm$0.02 & TICv8 \\ 
~~~W3 & WISE 3 Magnitude & 10.71$\pm$0.01 & 8.92$\pm$0.03 & TICv8 \\ 
~~~W4 & WISE 4 Magnitude & 8.8$^{b}$ & 8.76$\pm$0.02 & TICv8 \\ 
~~~T & TESS Magnitude & 13.966$\pm$0.007 & 11.737$\pm$0.007 & TICv8 \\ 
\sidehead{Spectroscopic Parameters}
~~~T$_{\rm{eff}}$  & Stellar Effective Temperature & 3214$\pm$69 & 3443$\pm$69 & This Work \\
~~~[Fe/H]  & Stellar Metallicity & 0.25$\pm$0.12 & -0.08$\pm$0.12 & This Work\\
~~~$\log$ g & Log Surface Gravity & 4.96$\pm$0.04 & 4.91$\pm$0.04 & This Work \\ 
\sidehead{Model Parameters$^{c}$}
~~~T$_{\rm{eff}}$  & Stellar Effective Temperature & $3168^{+39}_{-35}$ & 3366$^{+39}_{-41}$ & This Work \\
~~~[Fe/H]  & Stellar Metallicity & 0.19$\pm$0.09 & -0.02$^{+0.07}_{-0.03}$ & This Work \\
~~~$\log g$  & Log Surface Gravity & 4.96$\pm$0.03 & 4.92$\pm$0.03 & This Work \\ 
~~~$M_{*}$ & Stellar Mass ($M_{\odot}$) & 0.27$\pm$0.02 & 0.34$\pm$0.02 & This Work \\ 
~~~$R_{*}$ & Stellar Radius ($R_{\odot}$) & 0.287$\pm$0.008 & 0.335$\pm$0.009 & This Work \\ 
~~~$L_{*}$ & Stellar Luminosity ($L_{\odot}$) & 0.0075$\pm$0.0002 & 0.0130$^{+0.0004}_{-0.0005}$ & This Work \\ 
~~~$\rho_{*}$ & Stellar Density (cgs) & 16.3$^{+1.2}_{-1.1}$ & 12.68$^{+0.88}_{-0.85}$ & This Work \\ 
~~~Age & Stellar Age (Gyr) & 7.1$\pm$4.6 & 7.5$^{+4.2}_{-4.8}$ & This Work \\ 
\sidehead{Other Parameters}
~~~$v \sin i$  & Rotational Velocity (km s$^{-1})$ & $<$ 2 & $<$ 2 & This Work \\
~~~$\Delta$ RV  & Stellar Radial Velocity (km s$^{-1}$) & -4.1$\pm$0.1 & 0.0$\pm$0.1 & This Work  \\
~~~d  & Distance (pc) & 64.62$\pm$0.14 & 33.33$\pm$0.02 & This Work\\
\enddata
\tablenotetext{a}{Gaia EDR3 refers to \cite{gaia21}. TICv8 refers to \cite{stassun18}. 2MASS refers to \cite{cutri03}. APASS DR10 refers to \citep{henden18}.}
\tablenotetext{b}{TICv8 and WISE catalogs do not report an uncertainty for this measurement.}
\tablenotetext{c}{Adopted values.}

\normalsize
\end{deluxetable*}

\section{Analysis}\label{sec:analysis}

\begin{deluxetable*}{llll}
\tablecaption{Priors Used for Bayesian Model Fits \label{priortable}}
\tablehead{\colhead{~~~Parameter Name} &
\colhead{Prior} &
\colhead{Units} & \colhead{Description}
}
\startdata
\sidehead{\textbf{TOI-1696:}}
\sidehead{Orbital Parameters}
~~~P$_{b}$ & $\mathcal{N}^{a}(2.5007,0.1)$ & days & Period\\
~~~T$_{c}$ & $\mathcal{N}(2458816.697706,0.1)$ & BJD (days) & Transit Time \\ 
~~~e & 0 (Fixed) & ...  & Eccentricity \\
~~~$\omega$ & 90 (Fixed) & degrees  & Argument of Periastron \\
~~~$R_{p}/R_{*}$ & $\log\mathcal{N}(-2.211,1.0)$ & ... & Scaled Radius \\
~~~b & $\mathcal{U}^{b}(0.0, 1.0)$ & ...  & Impact Parameter \\
~~~K$_{b}$ & $\mathcal{U}(0.01,100)$ & m s$^{-1}$  & Velocity Semi-amplitude \\
\sidehead{Instrumental Parameters}
~~~$\gamma_{\rm{HPF}}$ & $\mathcal{U}(-100,100)$ & m s$^{-1}$ & Instrumental RV Offset \\
~~~$\dot{\gamma}_{\rm{HPF}}$ & $\mathcal{U}(-100, 100)$ & m s$^{-1}$ yr$^{-1}$  & RV Trend \\ 
~~~$\sigma_{\rm{HPF}}$ & $\mathcal{U}(0.01,100)$ & m s$^{-1}$  & RV Jitter \\
~~~$\sigma_{\rm{TESS}}$ & $\log\mathcal{N}(-7.58,2)$& ... & Photometric Jitter \\
~~~$\sigma_{\rm{ARCTIC}}$ & $\log\mathcal{N}(-11.68,2)$& ... & Photometric Jitter \\
~~~$\sigma_{\rm{RBO}}$ & $\log\mathcal{N}(-8.50,2)$& ... & Photometric Jitter \\
~~~$\gamma_{\rm{TESS}}$ & $\mathcal{N}(0.0,10.0)$& ... & Photometric Offset \\
~~~$\gamma_{\rm{ARCTIC}}$ & $\mathcal{N}(0.0,10.0)$& ... & Photometric Offset \\
~~~$\gamma_{\rm{RBO}}$ & $\mathcal{N}(0.0,10.0)$& ... & Photometric Offset \\
~~~$u_{\rm{TESS}}$ & $\mathcal{K}^{c} $ & ... & Quadratic Limb Darkening \\
~~~$u_{\rm{ARCTIC}}$ & $\mathcal{K}$& ... & Quadratic Limb Darkening \\
~~~$u_{\rm{RBO}}$ & $\mathcal{K} $ & ... & Quadratic Limb Darkening \\
~~~Dil$_{\rm{TESS}}$ & $\mathcal{U}(0,2)$ & ... & Dilution \\
~~~s & $\mathcal{U}(0,10)$ & ... & RBO Jitter Scale \\
\sidehead{\textbf{TOI-2136:}}
\sidehead{Orbital Parameters}
~~~P$_{b}$ & $\mathcal{N}(7.851866,0.1)$ & days & Period\\
~~~T$_{c}$ & $\mathcal{N}(2459017.704899,0.1)$ & BJD (days) & Transit Time \\ 
~~~e & 0 (Fixed) & ...  & Eccentricity \\
~~~$\omega$ & 90 (Fixed) & degrees  & Argument of Periastron \\
~~~$R_{p}/R_{*}$ & $\log\mathcal{N}(-2.71,1.0)$ & ... & Scaled Radius \\
~~~b & $\mathcal{U}(0.0, 1.0)$ & ...  & Impact Parameter \\
~~~K$_{b}$ & $\mathcal{U}(0.01,100)$ & m s$^{-1}$  & Velocity Semi-amplitude \\
\sidehead{Instrumental Parameters}
~~~$\gamma_{\rm{HPF}}$ & $\mathcal{U}(-100,100)$ & m s$^{-1}$ & Instrumental RV Offset \\
~~~$\dot{\gamma}_{\rm{HPF}}$ & $\mathcal{U}(-100, 100)$ & m s$^{-1}$ yr$^{-1}$  & RV Trend \\ 
~~~$\sigma_{\rm{HPF}}$ & $\mathcal{U}(0.01,100)$ & m s$^{-1}$  & RV Jitter \\
~~~$\sigma_{\rm{TESS}}$ & $\log\mathcal{N}(-7.94,2)$& ... & Photometric Jitter \\
~~~$\sigma_{\rm{ARCTIC}}$ & $\log\mathcal{N}(-6.21,2)$& ... & Photometric Jitter \\
~~~$\gamma_{\rm{TESS}}$ & $\mathcal{N}(0.0,10.0)$& ... & Photometric Offset \\
~~~$\gamma_{\rm{ARCTIC}}$ & $\mathcal{N}(0.0,10.0)$& ... & Photometric Offset \\
~~~$u_{\rm{TESS}}$ & $\mathcal{K} $ & ... & Quadratic Limb Darkening \\
~~~$u_{\rm{ARCTIC}}$ & $\mathcal{K}$& ... & Quadratic Limb Darkening \\
~~~Dil$_{\rm{TESS}}$ & $\mathcal{U}(0,2)$ & ... & Dilution \\
\enddata
\tablenotetext{a}{$\mathcal{N}$ is a normal prior with $\mathcal{N}$(mean,standard deviation)}
\tablenotetext{b}{$\mathcal{U}$ is a uniform prior with $\mathcal{U}$(lower,upper)}
\tablenotetext{c}{$\mathcal{K}$ is a reparametrization of a uniform prior for limb darkening, outlined in \citep{exoplanet:kipping13}}

\normalsize
\end{deluxetable*}

As detailed in Section \ref{sec:observations}, we have obtained photometry (TESS, RBO, ARCTIC) to constrain the transit events of TOI-1696 and TOI-2136, as well as RV data (HPF) to constrain the orbital parameters of each planetary system. Final parameter estimation is taken from a joint fit between the photometry and the RVs for both systems. In this joint fit, a few of the orbital parameters are shared: period, transit time, eccentricity, stellar mass, and inclination. Most of these parameters are quite well constrained prior to our joint modeling. Period and transit time are tightly constrained by the SPOC pipeline \citep{jenkins16}, the inclination must be close to 90$^{o}$ by necessity, and the stellar mass is estimated to $>$ 15$\%$ precision from our SED fits. A detailed summary of priors is visible in Table \ref{priortable}. Because most shared parameters are already tightly constrained, we do not expect a joint fit to be significantly better than individual transit and RV fits. Indeed, we did run individual RV and transit fits of this system, and found the estimated posteriors to have no significant difference from the joint fits. Nonetheless, we adopt a joint fit as our best fit, as it is the single model with the most complete description of each system.

\subsection{Transit Analysis}
\label{sec:transitanalysis}

Both systems' photometry were analyzed using the \textsf{exoplanet} software package \citep{exoplanet:joss}. First, the TESS photometry was downloaded using \textsf{lightkurve} \citep{lightkurve} for both targets. Data points flagged as poor quality during the SPOC pipeline were then discarded, and we median-normalized the lightcurves of both targets and centered them at 0. Then we imported our additional ARCTIC and RBO photometry and combined the datasets.

We used \textsf{exoplanet} to construct a physical transit model for each system. These models consisted of mean term for each instrument, 3 for TOI-1696, and 2 for TOI-2136, to account for any offsets. The same number of jitter terms were used to account for excess white noise in each instrument.

As mentioned in Section \ref{sec:rbo}, the RBO photometry exhibits a peculiar increase in scatter throughout the night. Unable to account for this by decorrelating with airmass, and with no additional explanations revealed in the night observing logs, we adopt a method in our model to increase the jitter of RBO data as the night goes on. This modified RBO jitter is given in equation \ref{modifiedjitter}. In this equation, $\sigma_{RBO}^{\prime}[i]$ represents the i$^{th}$ component of the vector of values that we add in quadrature to the RBO errorbars. $\sigma_{RBO}$ is the traditional jitter term analogous to those used in TESS and ARCTIC fits. t$_{i}$ is the time passed since the first RBO observation. t$_{tot,RBO}$ is the total duration of the RBO observations, and $s$ is the RBO jitter scale, a new free parameter to control how much the error bars should increase in time.

\begin{equation}
\label{modifiedjitter}
\centering
\sigma_{RBO}^{\prime}[i] = \sigma_{RBO}(1 + t_{i}/t_{tot,RBO})^{s}
\end{equation}

\begin{figure*}[] 
\centering
\includegraphics[width=\textwidth]{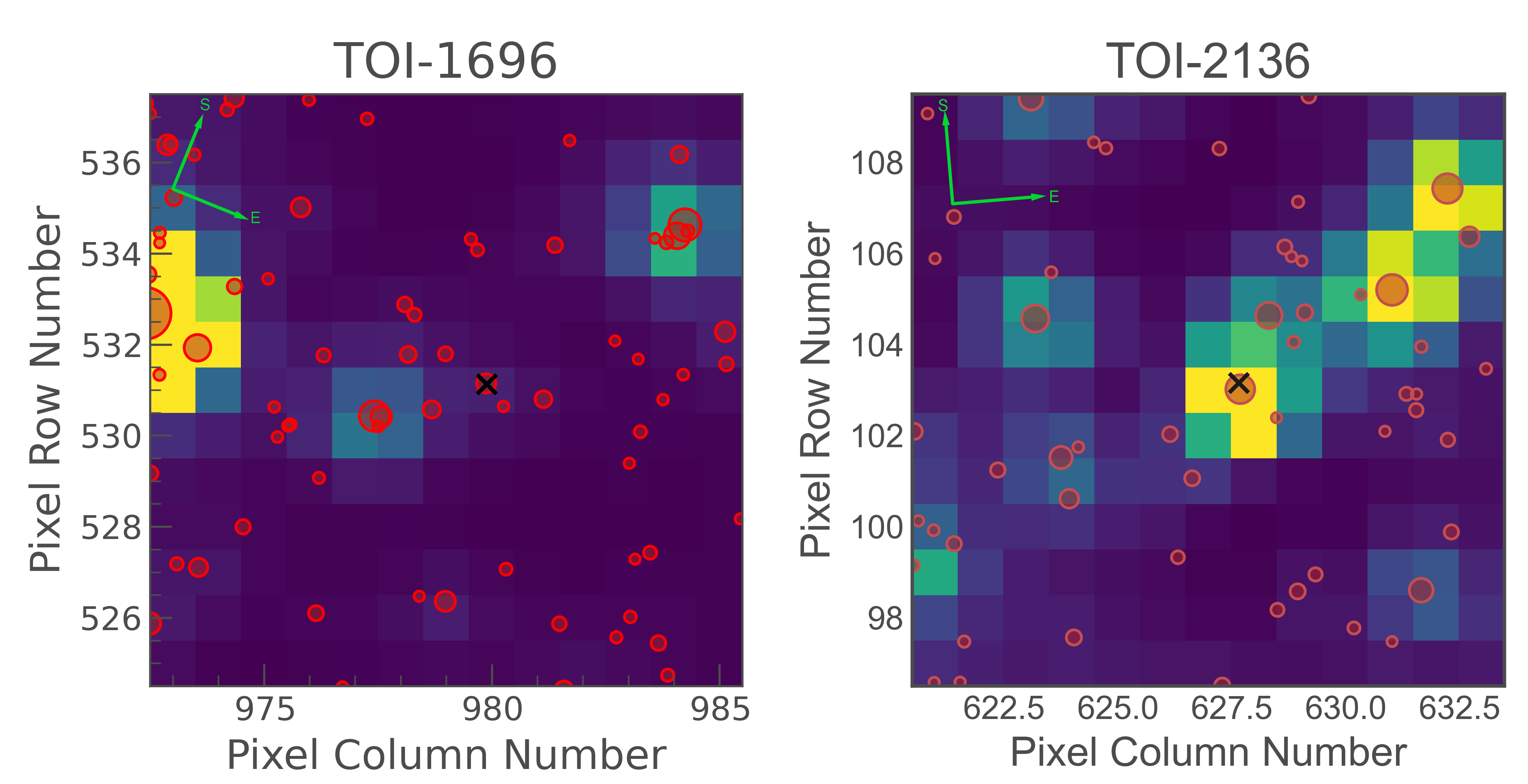}
\caption{TESS pixel plots of TOI-1696 and TOI-2136 made using the \textsf{eleanor} software package \citep{feinstein19}. Each square is a TESS pixel. The color of a pixel indicates the flux present. A black x near the center represents the TICv8 resolved position of the source. Red circles are Gaia resolved sources with Gaia magnitudes $<$ 19, and the size of the circle represents the brightness of the source. The fields for both stars have many possible sources of contamination. The left field is of Sector 19, the right field of Sector 26.} 
\label{fig:pixels}
\end{figure*}

The SPOC pipeline \citep{jenkins16} performs an automatic dilution adjustment on the PDCSAP flux of TESS lightcurves. This dilution increases the depths of transits to account for flux from nearby, adjacent stars, and is particularly important in crowded fields \citep{burt20}. We include pixel images of TOI-1696 and TOI-2136 in Figure \ref{fig:pixels}. We see from inspection that both fields are crowded, and that a dilution term might be important. TOI-1696 has an estimated contamination ratio of 0.1102, meaning that 11$\%$ of its flux is likely from nearby sources \citep{stassun19}. TOI-2136 has a contamination ratio of 0.1498. Both values warrant caution during analysis. Thus, we adopt as a part of our transit model a dilution term that floats between 0 and 2. A value of $<$ 1 suggests that contamination is still present, and additional correction is required. A value $>$ 1 suggests that the flux has been over-corrected for dilution by the SPOC pipeline. The model radius parameter is multiplied by the square root of this dilution term to allow an increase or decrease depending on dilution.

The transiting orbit model was generated using built-in \textsf{exoplanet} functions and the \texttt{starry} lightcurve package \citep{exoplanet:luger18}, which models the period, transit time, stellar radius, stellar mass, eccentricity, radius, and impact parameter to produce a simulated lightcurve. We adopt quadratic limb darkening terms to account for the change in flux that occurs when a planet approaches the limb of a star \citep{exoplanet:kipping13}.

A Gaussian Process (GP) model \citep{ambikasaran15}  was considered to account for excess correlated noise, as this is often done when analyzing the TESS photometry of even quiet stars \citep[e.g][]{191939}. However, both systems' photometry showed little evidence for coherent noise, and GP whitened results showed no signficant difference from models without GPs. Thus, in the pursuit of simplicity, we dispensed with any pre-fit whitening and fit transits to PDCSAP flux with no modification.

We chose to adopt fairly broad Gaussian priors with a width of 0.1 days for period and transit time to prevent any bias in our fits. Other free parameters had broad priors to reflect the wide array of possible values. A full list of the priors used is listed in Table \ref{priortable}. 

Due to the proximity of both systems to their host stars, and the estimated age of each system, we attempt to determine whether eccentric fits were reasonable by calculating the circularization time for both planets. This time is calculated using equation \ref{eqn:circularization}, as detailed in \citep{goldreich66}.

\begin{equation}
    \tau_{circ} = \frac{2PQ^{\prime}}{63\pi}\bigg(\frac{M_{p}}{M_{*}}\bigg)\bigg(\frac{a}{R_{p}}\bigg)^{5}
    \label{eqn:circularization}
\end{equation}

Here, P is the planet's orbital period, Q$^{\prime}$ is a friction coefficient, M$_{p}$ is the planet mass and M$_{*}$ is the stellar mass. R$_{p}$ is the planet radius, and a is the semi-major axis.

Because neither system has a well-constrained planet mass, we have opted to use the 3$\sigma$ upper limits of our mass estimates, since circularization time increases with planetary mass. The parameter $Q^{\prime}$ isn't known for either system, and must be estimated. This parameter represents the efficiency with which energy is lost due to tidal deformation. We adopt a similar approach to that used in \cite{waalkes21}, and attribute to each planet a $Q^{\prime}$ = 1 $\times$ $10^{4}$ since they are both in the radius regime of mini-Neptunes, though we caution that this is an assumption. 

For TOI-1696, we estimate a circularization timescale of 0.062 Gyr using the 3$\sigma$ upper limit of mass M$_{p}$ = 56.6 M$_{\oplus}$. Using the median estimate of the mass results in an even smaller timescale, 0.011 Gyr. We therefore conclude that eccentricity is unlikely to be present in this system, and can be fixed to 0.

On the other hand, for TOI-2136 we estimate a circularization timescale of 24.2 Gyr if we take its mass at a 3$\sigma$ upper limit of 15.0 $M_{\oplus}$. Even taking the median value of 4.70 $M_{\oplus}$ gives a circularization timescale of 7.49 Gyr, well within the uncertainties of our stellar age. We thus conclude that eccentric fits are perfectly feasible for TOI-2136, and must be considered in our final results.

\begin{figure*}[] 
\centering
\includegraphics[width=\textwidth]{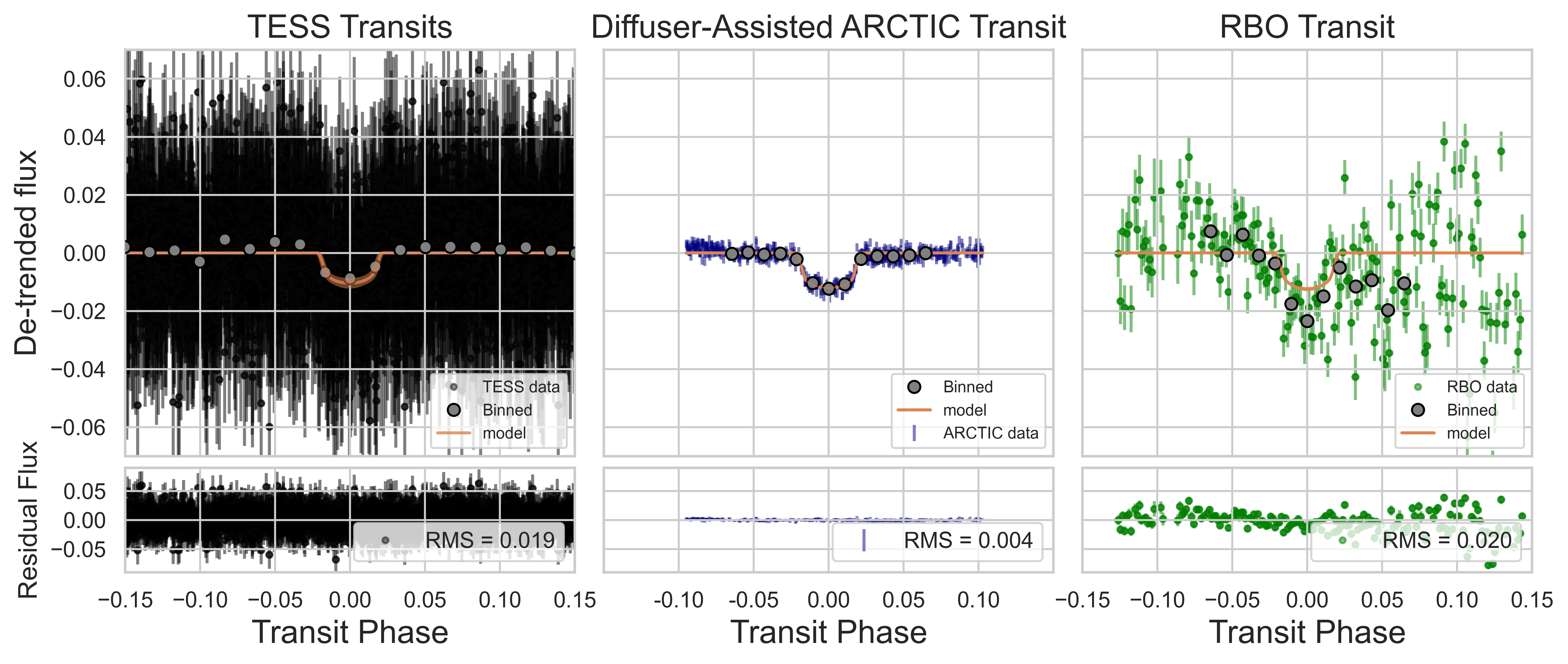}
\caption{Left: phase-folded TESS transits of TOI-1696, with binned points highlighted and the median lightcurve prediction overlaid. Middle: ARCTIC transit of TOI-1696 captured on 1 January 2021. Binned photometry and the median lightcurve prediction are overlaid. Right: RBO transit of TOI-1696 captured on 27 December 2020. Binned photometry and median lightcurve prediction are overlaid.} \label{fig:transitfit1696}
\end{figure*}

\begin{figure*}[] 
\centering
\includegraphics[width=\textwidth]{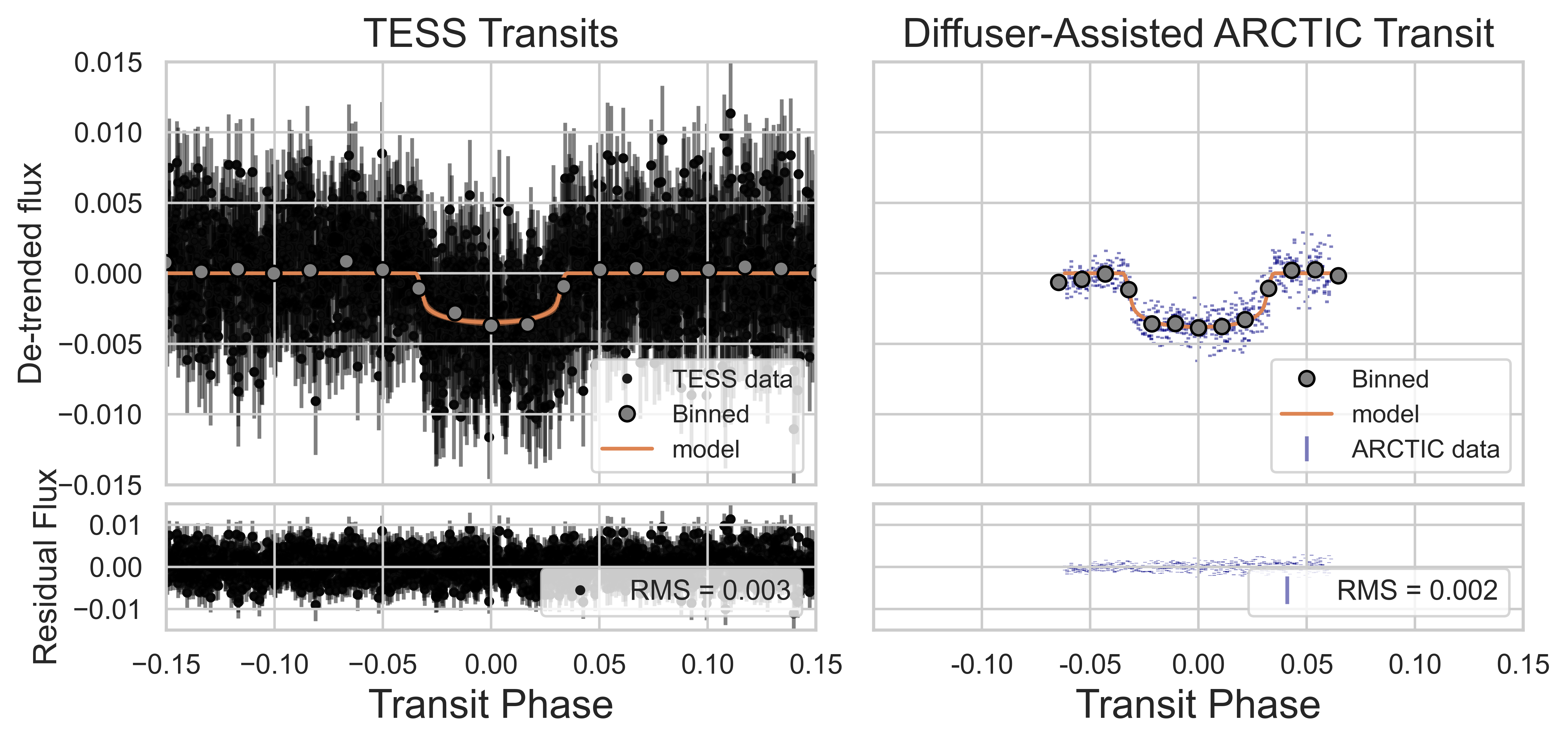}
\caption{Left: TESS photometry of TOI-2136 folded to the final estimated period for planet b. Binned points are plotted in addition to the median lightcurve prediction. Right: ARCTIC transit of TOI-2136 obtained on 5 October 2020. Binned photometry is plotted in gray in addition to the median lightcurve prediction in orange.} \label{fig:transitfit2136}
\end{figure*}

The total models for each system were then optimized using \texttt{scipy.optimize} to find a maximum a posteriori (MAP) fit to provide a starting point for posterior inference \citep{vertanen20}. We then ran a Markov Chain Monte Carlo (MCMC) sampler to explore the posterior space of each model parameter. We use the Hamiltonian Monte Carlo (HMC) algorithm with a No U-Turn Sampler (NUTS) for efficiency \citep{hoffman11}. We ran 10000 tuning steps and 10000 subsequent steps, and assessed convergence criteria using the Gelman-Rubin (G-R) statistic \citep{ford06}. The final transit fit of TOI-1696 is visible in Figure \ref{fig:transitfit1696}, and of TOI-2136 in Figure \ref{fig:transitfit2136}. The posterior parameters of each system are listed in Table \ref{paramtable}.

\subsection{Radial Velocity Analysis}

Both systems' radial velocities were analyzed independently using the \textsf{radvel} RV fitting software \citep{fulton18}, in addition to being used in a joint model with \textsf{exoplanet}. In both software packages the planet's RV signal is represented by a Keplerian orbit constrained by five planetary parameters: period, time of conjunction, RV semi-amplitude, eccentricity, and argument of periastron.

For TOI-1696, we detailed in Section \ref{sec:transitanalysis} that an eccentric fit is unlikely for this system, but an eccentric fit is plausible for TOI-2136. Nonetheless we run eccentric fits on the RVs for both systems, and perform a model comparison, seen in Table \ref{tab:modelcomparison}.

For both the RV-only fits and the joint-photometry fits, a jitter term is included to account for excess white noise, and a mean term is included to account for any systematic offset in the RVs, though with only one instrument this shouldn't be significantly different from 0. Due to the well constrained ephemeris in the TESS data, tight priors were placed on the period and time of conjunction of both systems. We used broad, uninformative priors for the remaining free parameters that are not constrained by photometry. A full list of our priors can be seen in Table \ref{priortable}. 

The RV only fits utilized the Powell optimization method \citep{powell98} to provide an initial starting guess for every parameter. We then ran an MCMC sampler in \textsf{radvel}, which utilizes the ensemble sampler outlined in \citet{foremanmackey13}, to explore the parameter space of the model. We used the G-R statistic again to assess convergence.

\begin{figure}[] 
\centering
\includegraphics[width=0.46\textwidth]{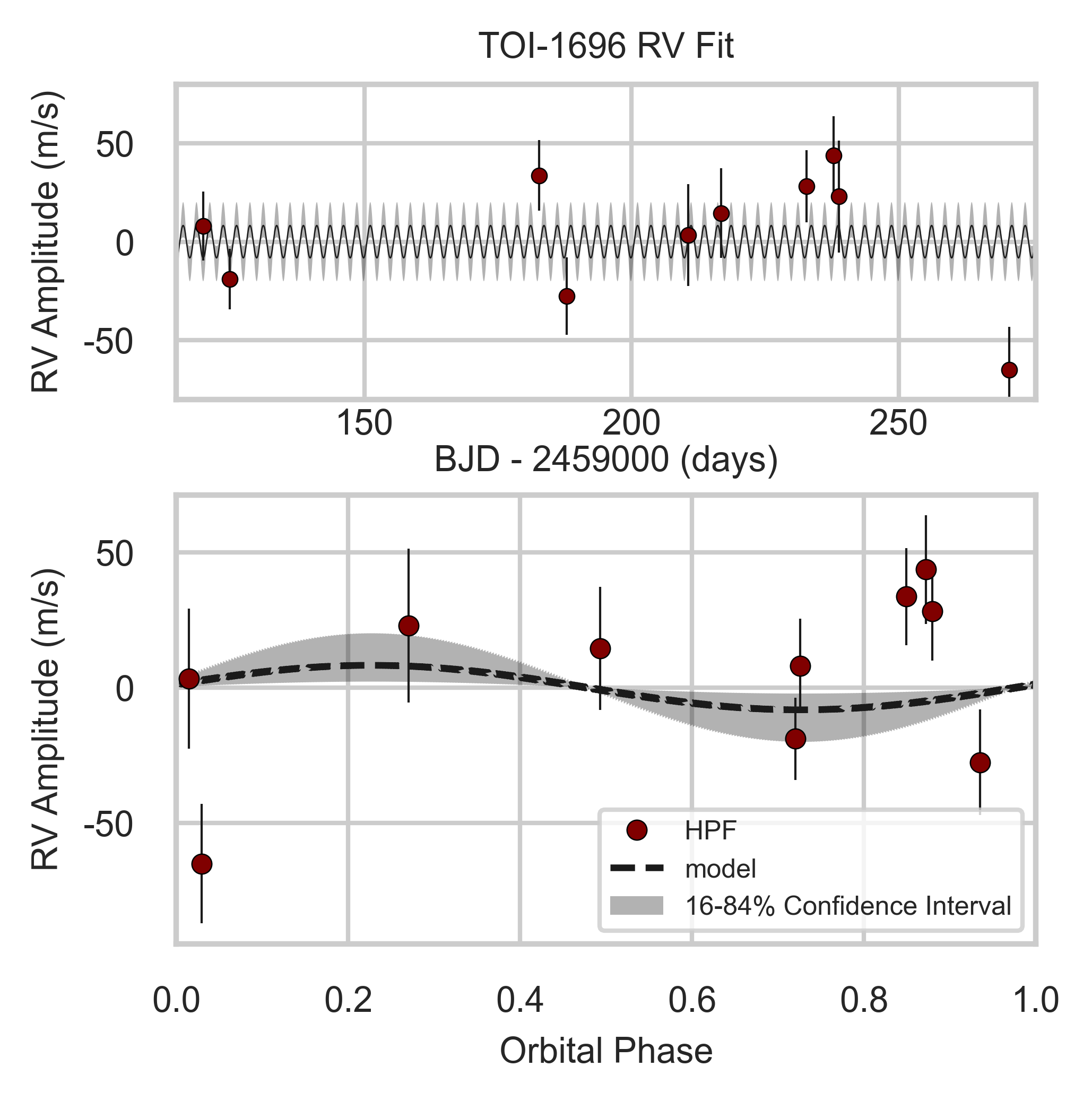}
\caption{Top: Total HPF RV timeseries of TOI-1696. Bottom: Final, phase-folded RV fit for TOI-1696b. Large uncertainties and a relatively small number of points only allow us to put an upper limit on the planet's amplitude. A 1$\sigma$ confidence interval is overlaid in gray.} \label{fig:rvfit1696}
\end{figure}

\begin{figure}[] 
\centering
\includegraphics[width=0.46\textwidth]{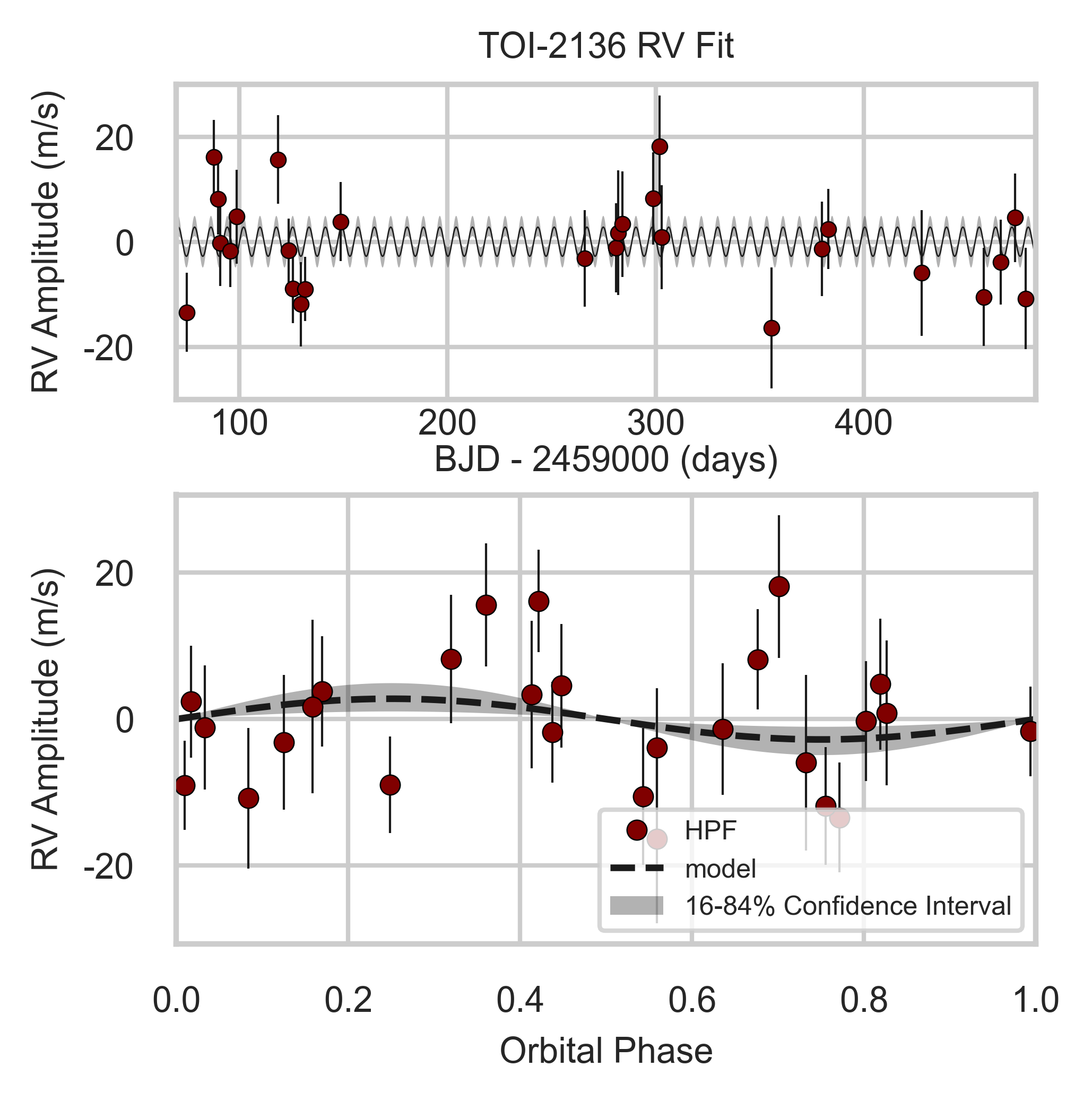}
\caption{Top: Total HPF RV timeseries of TOI-2136. Bottom: Final RV fit for TOI-2136b, folded to the estimated period of the planet. A 1$\sigma$ confidence interval is overlaid.} \label{fig:rvfit2136}
\end{figure}

The joint RV-photometry fits were performed in \textsf{exoplanet} in a nearly identical framework to the transit analysis described in Section \ref{sec:transitanalysis}, with the addition of the RV data and free parameters listed above. The final outputs from the joint fit are listed in Table \ref{paramtable}. The final RV fits can be seen for TOI-1696 in Figure \ref{fig:rvfit1696}, and TOI-2136 in Figure \ref{fig:rvfit2136}.

\begin{deluxetable}{ccccc}
\tablecaption{RV Model Comparisons$^{a}$ \label{tab:modelcomparison}}
\tablehead{\colhead{Fit}  &  \colhead{Free}   & \colhead{Number of}
& \colhead{BIC} & \colhead{RMS}\\
& \colhead{Parameters} & \colhead{Parameters} & & m s$^{-1}$}
\startdata
\sidehead{\textbf{TOI-1696:}}
Circular & $K$, $P$, $T_{c}$, & 6 & 111.24 & 31.36 \\
 & $\gamma$, $\sigma$, $dv/dt$ & & \\
Eccentric & $K$, $P$, $T_{c}$, $\gamma$, & 8 & 115.82 & 31.36 \\
 & $\sigma$, $dv/dt$, $e$, $\omega$ & & \\
\sidehead{\textbf{TOI-2136:}}
Circular & $K$, $P$, $T_{c}$, & 6 & 213.59 & 8.66  \\
 & $\gamma$, $\sigma$, $dv/dt$ &  &  & \\
Eccentric & $K$, $P$, $T_{c}$, $\gamma$, & 8 & 214.64 & 8.10 \\
 &  $\sigma$, $dv/dt$, $e$, $\omega$ & & \\
Circular + GP & $K$, $P$, $T_{c}$, $\gamma$, & 10 & 223.88 & 6.01 \\
 & $\sigma$, $dv/dt$, $\eta_{1}$, & & \\
  & $\eta_{2}$, $\eta_{3}$, $\eta_{4}$ & &\\
\enddata
\tablenotetext{a}{Model comparison was performed on RV-only fits. Transit parameters are well behaved, and were never significantly different between models. Furthermore, the primary model differences (eccentricity and GP) are most relevant for the RVs}
\end{deluxetable}

\subsubsection{RV Analysis of TOI-1696}

As detailed in Section \ref{sec:transitanalysis}, TOI-1696b is unlikely to have an eccentric orbit. Nonetheless, we allowed the eccentricity and argument of periastron to vary in some of our fits, and performed a model comparison to evaluate which model fits the data the best. These comparisons are visible in Table \ref{tab:modelcomparison}. We use the Bayesian Information Criterion \citep[BIC;][]{kass1995} to compare models. A ``Bayes Factor" is computed as the half the difference in BIC of the simpler model minus the more complex model. A Bayes Factor of $>$ 3.2 suggests a substantial preference for the more complex model. 

After analysis, the circular model is slightly preferred, though the difference in BIC values is too small to be considered statistically significant. Because the circular model has fewer free parameters, and because of our circularization arguments in Section \ref{sec:transitanalysis}, we adopt a circular fit as our best solution.

\subsubsection{RV Analysis of TOI-2136}

As mentioned in Section \ref{sec:transitanalysis}, TOI-2136b has a long circularization timescale, and we cannot rule out an eccentric orbit. Eccentric and circular fits were evaluated, and compared using their BIC. Our results in Table \ref{tab:modelcomparison} indicate that a more complex model cannot be justified. We go forward with a circular fit since it has fewer free parameters.

When we first chose TOI-2136 as a target, we made white-noise error estimates to determine the number of RVs that would be required to measure the mass of planet b to 3$\sigma$. Using the Mass-Radius relationship in \cite{kanodia19}, we estimated that the planet would have a semi-amplitude of 4 m s$^{-1}$  and that we would have a photon-noise single-measurement error of 6.5 m s$^{-1}$. Our estimated posterior amplitude (K$_{med}$ = 3.02 m s$^{-1}$) and median error ($\sigma_{median}$ = 7.87 m s$^{-1}$) suggest that both estimates were reasonable. However, our final results have a much less significant mass measurement at $K/\sigma_K<$ 2. This suggests three possible explanations: the planet's mass is significantly smaller than our median prediction, activity from the star is interfering with our mass measurements, or the existence of additional planets may be confounding our models.

We used a Generalized Lomb-Scargle periodogram (GLS; \citealt{zechmeister09}) to analyze several activity indicators (line width, Ca infrared triplet). Results suggest that this is a quiet star, though modestly strong peaks at $\sim$ 5, 19, and 60 days in the Ca infrared triplet and line width periodograms are suggestive of possible rotation periods. It is often possible for activity present in RVs, however, to have no clear signal in one or more activity indicators \citep{robertson14,lubin21}, and so we proceed forward with our investigation despite the lack of clear detections.

We have enough RVs for TOI-2136 such that we might utilize a GP without overfitting. GPs are commonly utilized to mitigate activity and improve model fits \citep{haywood14, lopezmorales16,dumusque19}. Our RV-only analysis utilized the robust Quasi-Periodic GP kernel \citep{fulton18}. The covariance matrix of this kernel is described in Equation \ref{eqn:qp}. It contains 4 hyperparameters to model the activity: $\eta_{1}$ is the amplitude of covariance, $\eta_{2}$ is the evolution timescale, $\eta_{3}$ is the recurrence time scale (usually the rotation period of the star), and $\eta_{4}$ is the structure parameter.

\begin{equation}
    \centering
    \label{eqn:qp}
    \sigma_{i,j} = \eta_{1}^{2} \exp\bigg(-\frac{|t_{i} - t_{j}|^{2}}{\eta_{2}^{2}} - \frac{\sin^{2}(\frac{\pi(t_{i}-t_{j})}{\eta_{3}})}{2\eta_{4}^{2}}\bigg)
\end{equation}

Our joint RV-Photometry-GP fits utilized the Rotation Term kernel in \textsf{exoplanet}, which is a combination of two Simple Harmonic Oscillator (SHO) kernels. The SHO kernel is fast and widely applicable to coherent stellar astrophysical noise sources \citep{exoplanet:foremanmackey18}. The Fourier transform of the SHO is known as the Power Spectral Density (PSD) and can be seen in equation \ref{eq:psd}.

\begin{equation}
\label{eq:psd}
\centering
S(\omega) = \sqrt{\frac{2}{\pi}} \frac{S_{0}\omega_{0}^{4}}{(\omega^{2}-\omega_{0}^{2})^{2} + \omega^{2}\omega_{0}^{2}/Q^{2}}
\end{equation}

The free parameters of this kernel are S$_{0}$, $\omega_{0}$, and Q. S$_{0}$ represents the power of the periodicity in Fourier space, $\omega_{0}$ is the angular frequency of the coherent noise, and Q is the quality factor.

The hyperparameters of the GPs were generally given wide priors when fit. $\eta_{1}$ was given a uniform prior from 0.1 m s$^{-1}$ to 50.0 m s$^{-1}$. The $\eta_{2}$ and $\eta_{4}$ parameters were given, broad, uniform log priors from -6 to 6. The rotation period prior, $\eta_{3}$, can often be restricted much better due to some independent measurement of the rotation period. We performed an autocorrelation function (ACF) rotation analysis on the TESS photometry using an internally developed pipeline (Holcomb et al. 2022, in prep), and found no signficant detection of a rotation period in either sector for TOI-2136. Additionally, we analyzed publicly available photometry from the All-Sky Automated Search for Supernovae (ASAS-SN; \citealt{shappee14,kochanek17}) and the Zwicky Transient Facility (ZTF; \citealt{masci19,bellm19}) using GLS periodograms. We found no significant signals corresponding to a rotation period in the ASAS-SN photometry, but we do note a strong signal at 85.6 days in the ZTF data. This value is consistent with a rotation period described in \cite{newton16}. Such a long rotation period is suggestive of a low amplitude activity signal. Regardless, analysis was performed with both tight priors around the purported stellar rotation period, and with loose, uniform priors from 1.0 - =200.0 days.

Both analyses resulted in posterior estimates nearly identical to those without the use of a GP. A comparison of the BIC of the RV-only GP fit, to other RV fits, is visible in Table \ref{tab:modelcomparison}.

The lower BIC (and thus higher loglikelihood) of the simpler, no GP, non-eccentric model, suggest that it is the preferred model, especially considering its smaller number of free parameters. We adopt this as our final fit.

\subsection{An Additional Planet Orbiting TOI-2136?}

With a GP unable to explain our low-significance mass measurement, we turn to the next possibility: an additional planet, or planets. We began with an internally developed pipeline that utilizes a Boxed-Least Squares (BLS; \citealt{kovacs02}) algorithm to search for an additional transiting planet in the TESS data. After subtracting our best-fit lightcurve model of TOI-2136b from the photometry, we ran the BLS analysis and noted no significant detections of additional transiting planets. 

To probe for smaller transiting exoplanets, we used the \texttt{Transit-Least Squares} (\texttt{TLS}) python package to check for additional signals with greater sensitivity \citep{hippke19}. This \texttt{TLS} method is more computationally intensive than the BLS, but adopts a more realistic transit shape, and is more sensitive to small-radius transiting exoplanets.

The \texttt{TLS} package initially recovers TOI-2136b with high significance. Masking the transits of the first planet, we ran the analysis again. The result was a forest of small peaks, with no single signal standing out as a clear candidate planet. \texttt{TLS} uses the Signal Detection Efficiency (SDE) to estimate significant periods. \cite{dressing15} suggest that an SDE $>$ 6 represents a conservative cutoff for a ``significant" signal, though others adopt higher values \citep{siverd12,livingston18}. The original transiting signal of TOI-2136 has an SDE of 17.0, indicating a highly significant detection. After masking planet b, the forest of peaks all fall under the SDE = 6 threshold, with the highest having SDE = 4.8. We conclude that none of these signals are transiting planets.

We therefore detect no additional transiting planets in the photometry of TOI-2136b. Analysis of RV residuals, after the fitting of a single planet, also returns a forest of low significance peaks, all below the 0.01$\%$ analytical false alarm probability (FAP) \citep{sturrock10}. This does not rule out an additional planet as an explanation for our low mass significance, especially considering that the transiting planet also falls below our significant threshold in RVs, but we find no definitive evidence of such a companion in RVs or photometry.

\begin{deluxetable*}{llcc}
\tablecaption{Derived Parameters for both systems \label{paramtable}}
\tablehead{\colhead{~~~Parameter} &
\colhead{Units} &
\colhead{TOI-1696b} & \colhead{TOI-2136b}
}
\startdata
\sidehead{Orbital Parameters:}
~~~Orbital Period $\dotfill$ & $P$ (days) $\dotfill$ & 2.50031$\pm$0.00002 & 7.85191$\pm$0.00004\\
~~~Eccentricity $\dotfill$ & $e$ $\dotfill$ & 0 (fixed) & 0 (fixed)\\
~~~Argument of Periastron $\dotfill$ & $\omega$ (degrees) $\dotfill$ & 90 (fixed) & 90 (fixed) \\
~~~RV Semi-Amplitude$^{a}$ $\dotfill$ & $K$ (m/s) $\dotfill$ &
$<$ 63.1 & $<$ 9.63 \\
~~~Systemic Offset $\dotfill$ & $\gamma$ (m/s) $\dotfill$ & 
9$\pm$13 & 0$\pm$1.8 \\
~~~RV trend $\dotfill$ & $dv/dt$ ($\unit{mm/s/yr}$)   &  0.00$\pm$0.03  & 0.00$\pm$0.03   \\ 
~~~RV jitter $\dotfill$ & $\sigma_{\mathrm{HPF}}$ (m/s) $\dotfill$ & 31$^{+14}_{-10}$ & 4.4$\pm$2.4\\
\sidehead{Transit Parameters:}
~~~Transit Midpoint $\dotfill$ & $T_C$ (BJD$\textsubscript{TDB}$) $\dotfill$ & 2458816.699$\pm$0.002 & 2459017.7039$\pm$0.0006 \\
~~~Scaled Radius $\dotfill$ & $R_{p}/R_{*}$ $\dotfill$ & 
0.104$\pm$0.002 & 0.058$\pm$0.001\\
~~~Scaled Semi-major Axis $\dotfill$ & $a/R_{*}$ $\dotfill$ & 17.7$\pm$0.6 & 35.0$\pm$1.3\\
~~~Impact Parameter $\dotfill$ & $b$ $\dotfill$ & 0.56$\pm$0.05  & 0.41$\pm$0.10 \\
~~~Orbital Inclination $\dotfill$ & $i$ (degrees) $\dotfill$ & 88.461$\pm$0.004 & 88.441$\pm$0.003\\
~~~Transit Duration $\dotfill$ & $T_{14}$ (days) $\dotfill$ &  0.0428$^{+0.0009}_{-0.0008}$ & 0.0693$\pm$0.0008\\
~~~Limb Darkening $\dotfill$ & $u_{1,\rm{TESS}}$, $u_{2,\rm{TESS}}$ $\dotfill$ & 0.5$^{+0.5}_{-0.3}$,0.1$\pm$0.4 & 0.3$^{+0.3}_{-0.2}$, 0.2$^{+0.4}_{-0.3}$ \\
 & $u_{1,\rm{ARCTIC}}$, $u_{2,\rm{ARCTIC}}$ $\dotfill$  & 0.4$\pm$0.3, 0.1$^{+0.4}_{-0.3}$ & 0.2$\pm$0.2, 0.3$^{+0.3}_{-0.4}$ \\
 & $u_{1,\rm{RBO}}$, $u_{2,\rm{RBO}}$ $\dotfill$ & 0.6$^{+0.6}_{-0.4}$,0.0$\pm$0.5 & ...\\
~~~Photometric Jitter $\dotfill$ & $\sigma_{\rm{TESS}}$ (ppm) $\dotfill$ & 104$^{+169}_{-79}$ & 31$^{+39}_{-21}$ \\
 & $\sigma_{\rm{ARCTIC}}$ (ppm) $\dotfill$  & 1030$\pm$91 & 929$\pm$28\\
 & $\sigma_{\rm{RBO}}$ (ppm) $\dotfill$ & 4300$\pm$1200 & ...\\
~~RBO Jitter Scale & $s$ $\dotfill$ & 3.1$\pm$0.6 & ...\\
~~~Photometric Mean $\dotfill$ & mean$_{\rm{TESS}}$ (ppm) $\dotfill$ & 65$\pm$188 & 32$\pm$19 \\
 & mean$_{\rm{ARCTIC}}$ (ppm)$\dotfill$  & 440$\pm$83 & 1980$\pm$59\\
 & mean$_{\rm{RBO}}$ (ppm) $\dotfill$ & 4449$\pm$1100 & ...\\
~~~Dilution$\dotfill$ & $D_{\mathrm{TESS}}$ $\dotfill$ & 0.86$\pm$0.14 & 0.92$\pm$0.07\\
\sidehead{Planetary Parameters:}
~~~Mass$^{a}$ $\dotfill$ & $M_{p}$ (M$_\oplus$) $\dotfill$ & $<$ 56.6 & $<$ 15.0 \\
~~~Radius$\dotfill$ & $R_{p}$  (R$_\oplus$) $\dotfill$& 3.24$\pm$0.12 & 2.09$\pm$0.08 \\
~~~Density$^{a}$ $\dotfill$ & $\rho_{p}$ (g/$\unit{cm^{3}}$)$\dotfill$ & $<$ 9.44 & $<$ 9.53 \\
~~~Semi-major Axis$\dotfill$ & $a$ (AU) $\dotfill$ & 0.0235$\pm$0.0006 & 0.054$\pm$0.001 \\
~~~Average Incident Flux$\dotfill$ & $\langle F \rangle$ ($\unit{W/m^2}$)$\dotfill$ & 18000$^{+972}_{-859}$ & 6000$^{+300}_{-260}$ \\
~~~Planetary Insolation$\dotfill$ & $S$ (S$_\oplus$)$\dotfill$ & 13.5$\pm$0.7 & 4.4$\pm$0.2 \\
~~~Equilibrium Temperature$^{b}$ $\dotfill$ & $T_{\mathrm{eq}}$ (K)$\dotfill$ & 533$\pm$7 & 403$\pm$5 \\
\enddata
\tablenotetext{a}{Represents a 3$\sigma$ (99.7$\%$) confidence upper limit}
\tablenotetext{b}{Estimated assuming an albedo of 0}

\normalsize
\end{deluxetable*}

\section{Discussion}\label{sec:discussion}

One of the most important open questions in the field of exoplanet astronomy is how planets retain or lose their atmospheres. This is particularly important when studying planets around M dwarf hosts. Not only are these stars the most abundant stellar type in the Galaxy \citep{henry18}, but their extreme UV environments likely play a significant role in sculpting their planets' atmospheres.

Mini-Neptunes present an ideal environment with which to study atmospheres of exoplanets, especially close-in ones that have excellent prospects for transmission spectroscopy. Their overall bulk densities suggest that these planets possess an atmosphere potentially dominated by a large H/He envelope.  Such planets' atmospheres are easier to study, and their abundance means that we have a large number of possible systems to choose from. Mass estimates, too, become important in this regime, as there are a number of different possibilities for sub-Neptune compositions that are not well understood \citep{zeng19,bitsch21,bean21}. 

While we have only managed to place upper limits on the masses of both TOI-1696 and TOI-2136, we have measured their radii to high precision (Table \ref{paramtable}). While we cannot claim to have detected a low density (and therefore large atmosphere), we can claim that it is quite unlikely for these planets not to have an extended atmosphere: for either to be primarily terrestrial, these planets would have to be truly unique in exoplanet parameter space, as any detections of such massive terrestrial planets so far have been erroneous (i.e. \citealt{rajpaul17}). Thus, we will go forward under the assumption that both TOI-1696b and TOI-2136b are at the very least not terrestrial.

We have calculated the Transmission Spectroscopy Metric (TSM; \citealt{kempton18}) for both targets: 89.8 and 92.0, respectively. The TSM is a metric developed to rate the value of a target's amenability to atmospheric follow-up using the JWST. Both values are at or above the suggested cutoff for transmission spectroscopy follow-up.

\subsection{TOI-1696}
\label{disc:1696}

With the combination of TESS photometry and ground-based photometric follow-up, we were able to put a tight constraint on TOI-1696b's radius. With an estimated radius of 3.24$\pm$0.12 R$_{\oplus}$, TOI-1696b falls into a surprising region of radius-stellar effective temperature parameter space. As early as the \kepler\ mission, a relationship between planet size and stellar temperature was observed: the occurrence rate of smaller, rocky planets increases significantly for cooler stars, while the presence of larger planets (R$_{p}$ $>$ 2.5 $R_{\oplus}$) falls off appreciably \citep{dressing15}. Later studies confirmed the veracity of this and noted that planets with radii above 2.8 R$_{\oplus}$ are particularly rare \citep{mulders15}. \kepler\ did not discover a large number of planets orbiting M dwarfs due to the nature of its mission: most observed M dwarfs were too far away and therefore dim for detailed study. It is only with the recent advent of TESS and its all-sky survey of nearby stars that we have been able to study the population of short period planets transiting these cooler stars \citep{ballard19}.

\begin{figure}[] 
\centering
\includegraphics[width=0.46\textwidth]{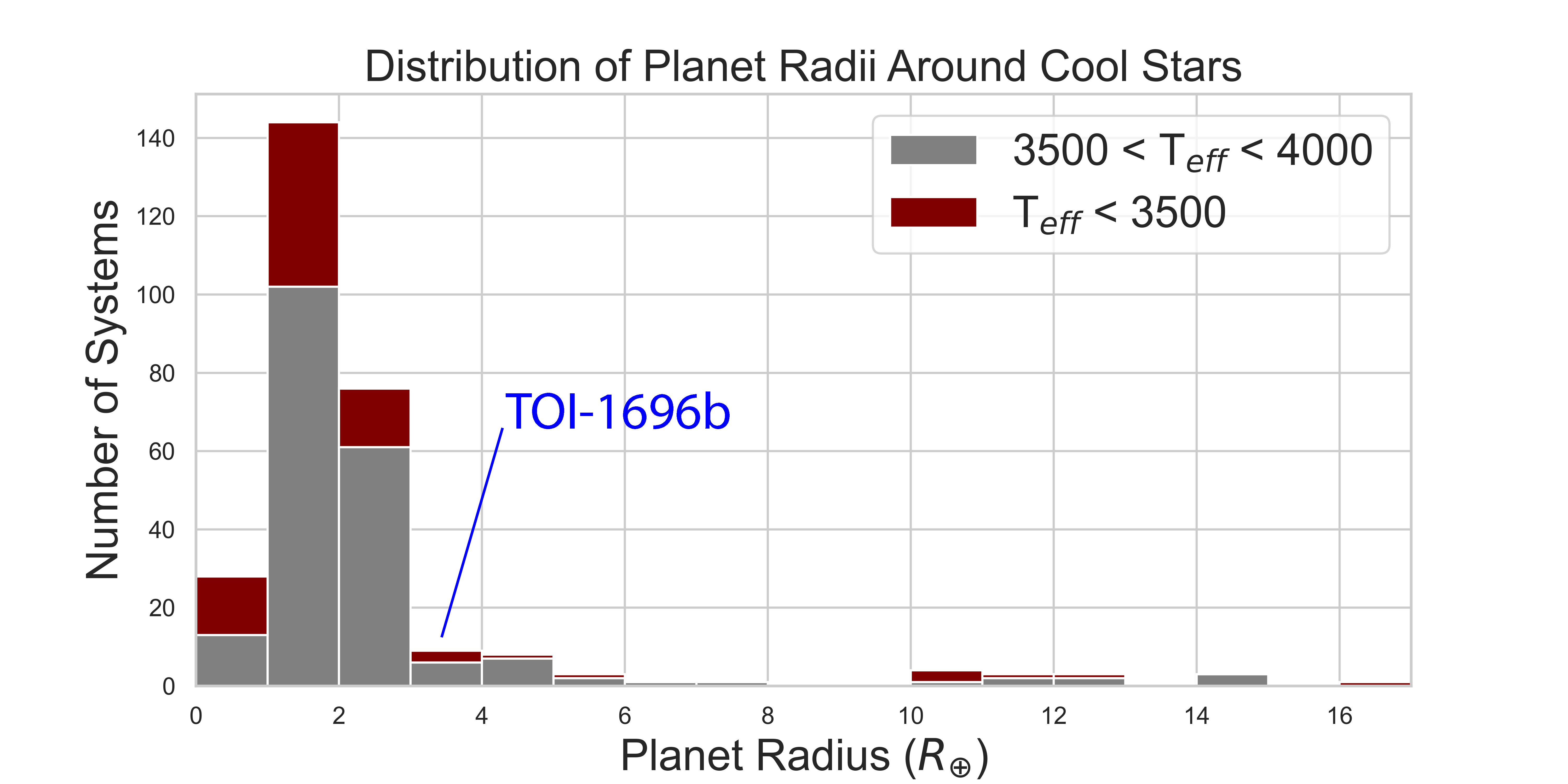}
\caption{Stacked histogram of the radii of known planetary systems around cool stars, taken from the \textsf{NASA Exoplanet Archive} \citep{akesone13}. The total sample is of all planets orbiting stars cooler than 4000 K. Notable is the relative paucity of giant planets around stars in this temperature regime. The occurrence rate is inversely proportional to $T_{\rm{eff}}$ \citep{dressing15}, and we further show a subset of this sample with $T_{\rm{eff}}$ $<$ 3500 K in red. TOI-1696b, with a radius of 3.24 $R_{\oplus}$ is unusually large for a star with $T_{\rm{eff}}$ = 3168 K, and its placement in the histogram is indicated in blue.} \label{fig:radiushist}
\end{figure}

\begin{figure*}[] 
\centering
\includegraphics[width=1.1\textwidth]{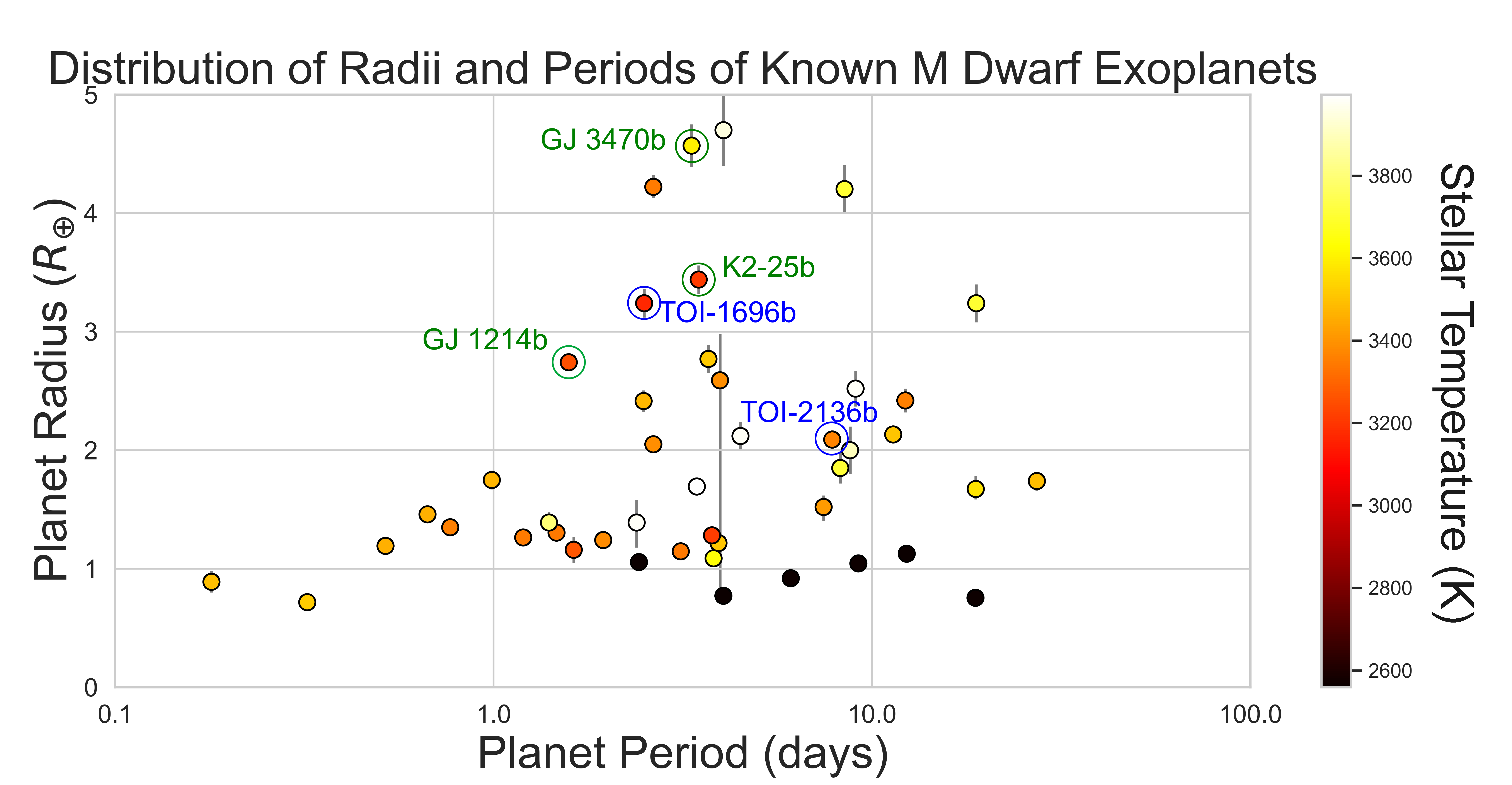}
\caption{Distribution of planets orbiting M dwarfs (T$_{\rm{eff}}$ $<$ 4000 K) in period-radius space taken from the \textit{NASA Exoplanet Archive} on 8 February 2022. To emphasize the fact that we are adding additional value by constraining the mass of TOI-1696b and TOI-2136b, we include only systems with at least upper limits on their mass$^{1}$. The most recently uploaded parameters are used. The planets studied in the paper are highlighted with blue circles, while a few notable systems are highlighted with green circles. TOI-1696b has a much larger radius than is typical considering its period, and the only stars with larger radii orbit hotter stars. Note also that with the exception of the TRAPPIST-1 planets at the bottom of the plot, TOI-1696b orbits among the coolest stars. TOI-2136b falls into a more common region of period-radius space, but has other attractive features.\\
\footnotesize
1. K2-25b does not yet have a mass uploaded to the NASA exoplanet archive, but we add it manually because of its similar nature to TOI-1696b \citep{stefansson20c}.} \label{fig:R-P}
\end{figure*}

In particular, there is more than an order of magnitude fewer exoplanets discovered around M dwarfs than there is around FGK dwarfs \citep{berger18}, despite the abunduance of M dwarf stars. As a result, phenemona such as the exoplanet radius valley \citep{fulton17,cloutier20, vaneylen21} are difficult to discern when looking at M dwarfs alone. Thus, there is great value in validating exoplanets orbiting M dwarfs, especially cooler ones where large exoplanets become very rare. In Figure \ref{fig:radiushist} it is clear that larger exoplanets are unusual around cooler stars, and that larger radii exoplanets become even more sparse the cooler the star. Only 10 currently confirmed planets orbit stars with T$_{\rm{eff}}$ $<$ 3500 K while also having a radius $>$ 2.8 R$_{\oplus}$ \citep{konopacky10,leggett14,artigau15,fontanive20,bakos20,stefansson20c,castrogonzalez20,wells21,parviainen21,zhang21}. Further, only 5 of these are close to their star, with P$_{orb}$ $<$ 100 days. An illustration of TOI-1696b's strange position in period-radius space is visible in Figure \ref{fig:R-P}. By studying a planet on the edge of M dwarf radius parameter space, we position ourselves to better answer questions about formation processes around cool stars, and to examine any dependencies or correlations with other physical parameters (i.e. metallicity). In particular, we expect comparisons between TOI-1696b and other mini-Neptunes in more common regions of paramater space to highlight exactly what qualities make larger mini-Neptunes around cool stars unlikely.

Additionally, for sub-Neptunes, there is some degeneracy between the compositions of planets between 2 - 4 $R_{\oplus}$. In particular, similar masses in this range can be explained either by gaseous planets with rocky cores and large envelopes of H/He, or ``water worlds" with large envelopes of H$_{2}$O fluid/ice, in addition to rock and gas (e.g. \citealt{zeng19}). Whether or not a sub-Neptune's atmosphere has lighter elements (H/He), or heavier (H$_{2}$O/CO$_{2}$) can have important implications for its formation and the history of the protoplanetary disk of the system \citep{valencia13,petigura17,anderson17}. Such information can even be used to infer the existence of large companions on much wider orbits \citep{bitsch21}.

TOI-1696b also falls near the ``Neptune desert," a region of parameter space where Neptune-sized objects become very rare \citep{mazeh16}. Several possible explanations for this ``Neptune desert," exist. For example, \cite{matsokos16} suggest that this feature is a natural result of tidal circularization as high eccentricity Neptunes exchange angular momentum with their host star. On the other hand, \cite{owen18} suggest that photoevaporation of short period sub-Jovians is the explanation for this feature. Detailed characterization of systems in and near this desert is required to break degeneracies between these explanations. 

\begin{figure*}[] 
\centering
\includegraphics[width=\textwidth]{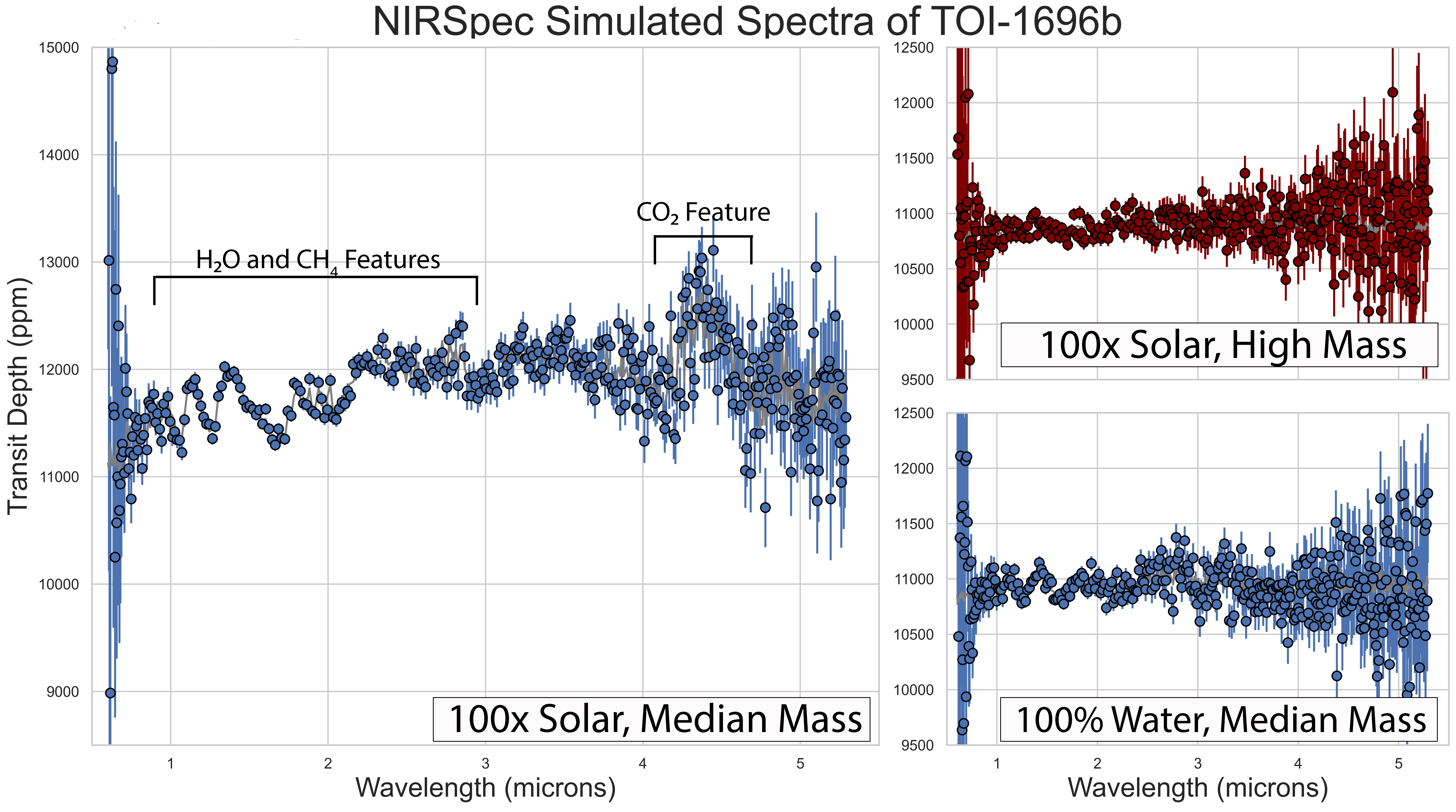}
\caption{Simulations of JWST observations of TOI-1696b using the NIRSpec instrument. Gray, predicted spectra are visible under the simulated data points. Data points are simulated with a 20 ppm systematic error noise floor. Simulations where the median planet mass (9.98} $M_{\oplus}$) was used are colored in blue, while simulations where the 3$\sigma$ upper limit (56.6 $M_{\oplus}$) were used are colored in red. Left: We see that a large H/He envelope with 100x solar metallicity has clearly resolvable water, carbon dioxide, and methane features. Right: Simulations of the massive H/He dominated planet, with 100x solar metallicity, above, and the less massive, but water dominated planet, below. Both make similar looking predictions, indicating the value that improving the mass measurement of TOI-1696b would add in the future. \label{fig:jwst_1696}
\end{figure*}

Transmission spectroscopy is one approach that can be used to address these issues, and TOI-1696b is a promising candidate for atmospheric study. Following \cite{kempton18}, we calculate a Transmission Spectroscopy Metric (TSM) of 89.8 using the median planet mass. JWST will soon be taking spectra of transiting exoplanets. In Figure \ref{fig:jwst_1696} we simulate several possibilities for atmospheric observations of TOI-1696b on JWST. 
We use \texttt{Exo-Transmit} \citep{exotransmit.reference} to create atmospheric models for TOI-1696b assuming a 100$\times$ Solar metallicity composition using its median mass of 9.98 M$_{\oplus}$ and for its upper 3$\sigma$ mass of 56.6 M$_{\oplus}$. We also created a ``steam'' atmosphere comprised of 100\% water for comparison. These models assume chemical equilibrium, are cloud-free and generated with an isothermal P-T profile with a planetary equilibrium temperature of 500 K. Using \texttt{PandExo} \citep{pandexo.reference}, we simulate expected JWST NIRSpec/Prism observations from 0.6 to 5 microns. In general, TOI-1696b is an excellent candidate for NIRSpec observations \citep{bagnasco07} assuming the median mass limit of the planet. We will be able to identify carbon dioxide, water, and methane in its atmosphere to better than 3$\sigma$ confidence with just 1 transit, and better than 5$\sigma$ with 2 transits. Resolving these features will allow us to measure the metallicity of the planet as well as the C/O and C/H ratios, enabling us to constrain the disk environment where this planet initially formed \citep[e.g.][]{oberg.co.ratio,moses.co.ratio}. At 500 K, TOI-1696b also lies in the interesting transition regime between ammonia and nitrogen dominated chemistries. Characterizing its atmospheric composition (and potentially detecting ammonia) would provide the first observations into the nitrogen-chemistries at play \citep{moses11}.

According to recent studies from \citet{yu.hazes,dymont.hazes}, the presence of aerosols (clouds or hazes) appear to be ubiquitous at this temperature regime. Therefore, it is probable that TOI-1696b also possesses an aerosol layer which could mute its absorption features in its spectrum. To test what we could observe with JWST, we add a simple (gray opacity, wavelength independent) aerosol layer at various pressure levels in our 100x Solar metallicity model (assuming median mass). We find that even at pressures of 0.1 mbars \citep[comparable to GJ 1214b][]{kreidberg.gj1214b}, we will still detect the presence of water, methane, and carbon dioxide. Moreover, it is predicted that moving to longer wavelengths should diminish the effect small haze particles play on a planet's transmission spectrum \citep{hu.and.kawashima.hazes}. We therefore do not expect the presence of aerosols to significantly hinder JWST observations of TOI-1696b. Instead, it is possible that these observations will in turn allow for a more detailed studies into the haze layer that we expect to be present in the planet's atmosphere.

However, we were only able to put an upper limit on the mass of this system, which can complicate the analysis of spectra obtained with JWST \citep{batlha19}. Simulations adopting the 3$\sigma$ upper mass limit of 56.6 $M_{\oplus}$ were also run for NIRSpec, pictured in Figure \ref{fig:jwst_1696}. Far fewer atmospheric features are discernible in such a situation, due mainly to the decrease in scale height associated with a larger planet mass. 

We anticipate TOI-1696b to be a system that receives much study from an atmospheric perspective in the future: its features are attractive from a wide number of scientific angles. Future RV measurements to better constrain the mass of this system would be a natural next step to understanding its composition and formation history. While the faintness of the star makes precision RVs challenging, higher precision could theoretically be obtained on future instruments for thirty-meter class telescopes. For example, using a rudimentary estimate of the performance of the Multi-Object Diffraction-limited High-resolution Infrared Spectrograph (MODHIS; \citealt{mawet19}) proposed for the Thirty-Meter Telescope (TMT), a 650 s exposure of TOI-1696 has an estimated photon-limited RV precision of $\sim$ 6 m s$^{-1}$. Considering that planet b has an estimated semi-amplitude of 11.7 m s$^{-1}$ from the mass-radius relationship in \cite{chen17}, we expect a more precise mass measurement for this system is well within reach of such an instrument.

\subsection{TOI-2136}
\label{disc:2136}

Using TESS photometry and HPF RVs, we were able to constrain TOI-2136b's radius to 2.09$\pm$0.08 R$_{\oplus}$, and its mass to $<$ 15.0 M$_{\oplus}$. This allowed us to constrain its density to $<$ 9.53 g/cm$^{3}$. We note, however, that MCMC chains have a median of 2.5 g/cm$^{3}$, which is consistent with a planet that is significantly less dense than Earth and other rocky exoplanets, and which has a significant gaseous envelope. In addition to our discussion of the infeasability of a $>$ 2 R$_{\oplus}$ planet being terrestrial in Section \ref{sec:discussion}, we proceed under the assumption that TOI-2136b has at least some sizable gaseous envelope.

TOI-2136b has potential as an exoplanet with detectable biosignatures. \cite{seager13a} first introduced the concept of a Cold-Haber World, where microbes live in liquid water environments beneath a large gaseous envelope. If such a gaseous envelope contained H$_{2}$ and even small amounts of N$_{2}$, organisms could potentially capture the energy used in the ``Haber" process, where H$_{2}$ and N$_{2}$ are combined exothermically to create NH$_{3}$, which was first proposed as a potential biosignature for H$_{2}$-rich worlds in \citet{seager13b}. NH$_{3}$ has several attractive features: its creation from the aforementioned exothermic process is energetically favorable; NH$_{3}$ is destroyed by photolysis, meaning that sustained production would be required to register a detection; and NH$_{3}$ is only produced abiotically in limited pressure-temperature regimes.

In fact, \cite{phillips21} established several criteria that make exoplanet candidates attractive for NH$_{3}$ biosignature detections. First, exoplanets with radii $>$ 1.75 R$_{\oplus}$ are best to ensure a gaseous envelope, but exoplanets with radii $<$ 3.4 R$_{\oplus}$ are best to ensure the pressure isn't great enough to produce abiotic NH$_{3}$. Next, a T$_{eq}$ $<$ 450 K allows liquid water to exist up to 1000 bar. The existence of liquid water is thought to be a necessary ingredient for life. Finally, systems with d $<$ 50 pc ensure adequate flux from star and planet for purposes of actually observing the biosignature with JWST. Our posterior results in Tables \ref{stellartable} and \ref{paramtable} place TOI-2136b comfortably within this regime. 

\begin{figure*}[] 
\centering
\includegraphics[width=\textwidth]{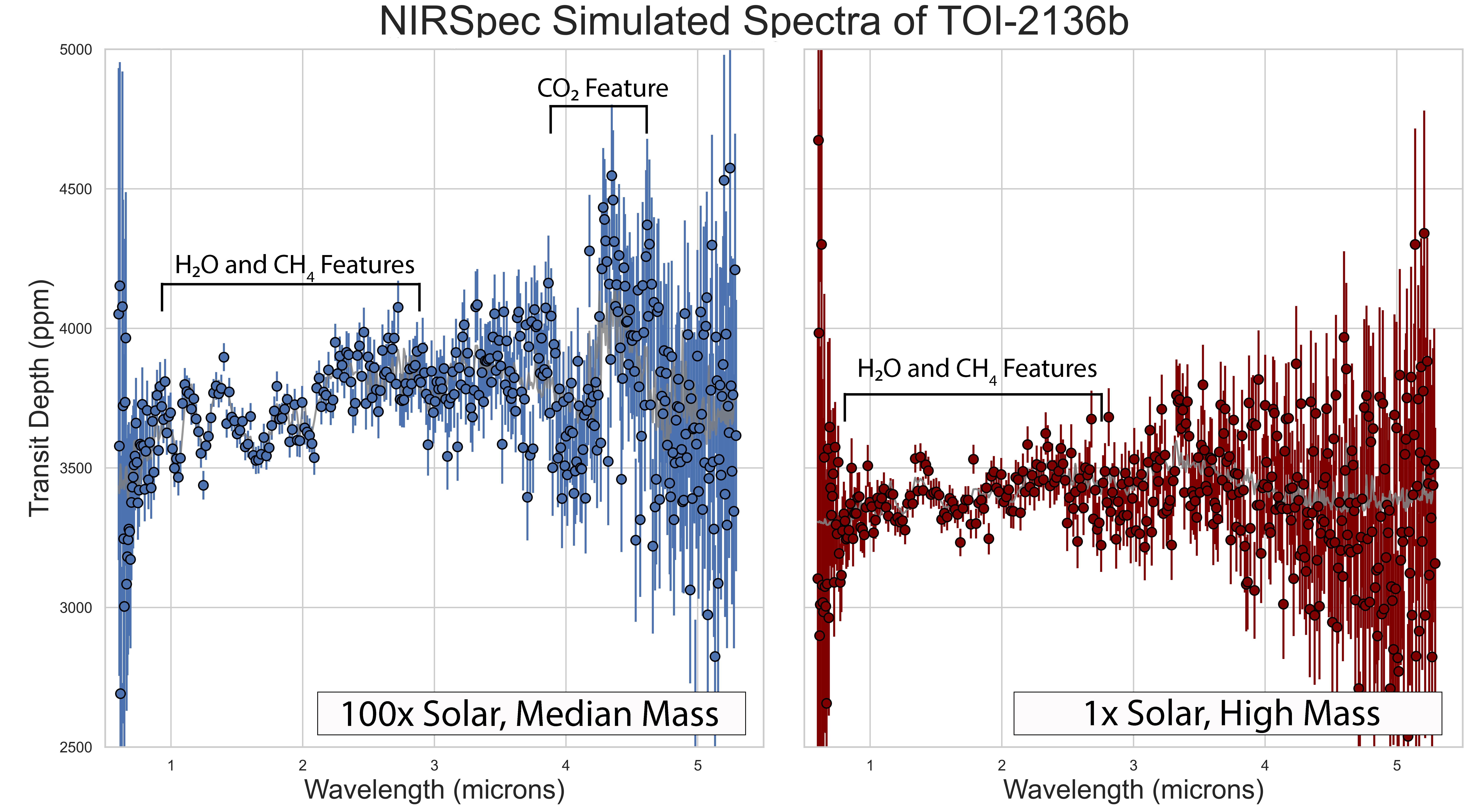}
\caption{Simulations of TOI-2136b with NIRSpec. The gray, predicted spectra are visible under the simulated data points. Left: simulated spectrum of the low mass planet with 100 times solar metallicity. Spectral features are clearly resolvable, especially water and methane features. Right: the high mass planet with solar-like metallicity in its atmosphere. Our mass limits constrain the planet's atmosphere enough such that both high and low mass scenarios have recoverable features.} \label{fig:jwst_2136}
\end{figure*}

We simulate NIRSpec observations of TOI-2136b using the same methods described in Section \ref{disc:1696}. Results of different simulated planet masses and metallicities are visible in Figure \ref{fig:jwst_2136}. Our tighter mass constraint of TOI-2136b allows us to resolve spectral features when using the median planet mass 4.64 M$_{\oplus}$ or when using the 3$\sigma$ upper limit of 15.0 M$_{\oplus}$ assuming a 1x Solar metallicity composition. We expect to recover better than 3$\sigma$ detections of water and methane features with just one transit, and better than 5$\sigma$ with two. Our simulations do not include an NH$_{3}$ term, as TOI-2136b is not expected to produce abiotic NH$_{3}$. \citet{phillips21} simulated NH$_{3}$ features for a system similar to TOI-2136b: TOI-270c (M$_{*}$ = 0.386; R$_{p}$ = 2.35 R$_{\oplus}$; P = 5.66 days; \citet{gunther19}). The TOI-270c system ranks highest in their metric for biosignature detection, and its simulated NH$_{3}$ features are recoverable in a small number of transits. This suggests that TOI-2136b will also rank very highly in future searches for atmospheric biosignatures with JWST.

We also note that during the submission of this manuscript, two additional studies were announced constraining the mass of TOI-2136b \citep{kawauchi22,gan22}. Our results are consistent with both estimates, though we note that our median planet mass of 4.64 M$_{\oplus}$ is more consistent with the analysis in \cite{kawauchi22}. The likelihood of a future analysis utilizing all three datasets only improves the attractiveness of TOI-2136b from an atmospheric study perspective.

\section{Summary}\label{sec:summary}

We refine the measured planetary, orbital, and stellar parameters of two TOI planet candidates, TOI-1696.01 and TOI-2136.01, and validate their planetary nature, using a combination of ground based photometry, high resolution adaptive optics imaging, and radial velocity measurements. 

Using ground-based photometry in coordination with TESS, we measure TOI-1696b's radius as 3.24 $\pm$ 0.12 R$_{\oplus}$, and TOI-2136b's radius as 2.09 $\pm$ 0.08 R$_{\oplus}$.

Using the near-IR Habitable-Zone Planet Finder, we are able to put upper limits on the masses of both transiting planets. We constrain TOI-1696b's mass to $<$ 56.6 M$_{\oplus}$, and TOI-2136b's mass to $<$ 15.0 M$_{\oplus}$, with 97.7$\%$ confidence. 

Both systems have high potential for future atmospheric studies, and detailed spectra of either system could answer important scientific questions. We encourage the community to continue observations on future instruments.

\section{Acknowledgements}

This paper includes data collected by the TESS mission. Funding for the TESS mission is provided by the NASA's Science Mission Directorate.

The Hobby-Eberly Telescope (HET) is a joint project of the University of Texas at Austin, the Pennsylvania State University, Ludwig-Maximilians-Universität München, and Georg-August-Universität Göttingen. The HET is named in honor of its principal benefactors, William P. Hobby and Robert E. Eberly.

The authors thank the HET Resident Astronomers for executing the observations included in this manuscript.

We would like to acknowledge that the HET is built on Indigenous land. Moreover, we would like to acknowledge and pay our respects to the Carrizo $\&$ Comecrudo, Coahuiltecan, Caddo, Tonkawa, Comanche, Lipan Apache, Alabama-Coushatta, Kickapoo, Tigua Pueblo, and all the American Indian and Indigenous Peoples and communities who have been or have become a part of these lands and territories in Texas, here on Turtle Island.

These results are based on observations obtained with the Habitable-zone Planet Finder Spectrograph on the HET. The HPF team was supported by NSF grants AST-1006676, AST-1126413, AST-1310885, AST-1517592, AST-1310875, AST-1910954, AST-1907622, AST-1909506, ATI 2009889, ATI-2009982, AST-2108512 and the NASA Astrobiology Institute (NNA09DA76A) in the pursuit of precision radial velocities in the NIR. The HPF team was also supported by the Heising-Simons Foundation via grant 2017-0494.


Based on observations at Kitt Peak National Observatory, NSF's National Optical-Infrared Astronomy Research Laboratory (NOIRLab Prop. ID: 2021A-0387; PI: Arvind Gupta), which is operated by the Association of Universities for Research in Astronomy (AURA) under a cooperative agreement with the National Science Foundation.


Observations in the paper made use of the NN-EXPLORE Exoplanet and Stellar Speckle Imager (NESSI). NESSI was funded by the NASA Exoplanet Exploration Program and the NASA Ames Research Center. NESSI was built at the Ames Research Center by Steve B. Howell, Nic Scott, Elliott P. Horch, and Emmett Quigley.

This work was supported in part by the National Science Foundation via grant AST-1910954 and AST-1907622.

This work was partially supported by funding from the Center for Exoplanets and Habitable Worlds.
The Center for Exoplanets and Habitable Worlds is supported by the Pennsylvania State University and the Eberly College of Science.

This research was supported in part by a Seed Grant award from the Institute for Computational and Data Sciences at the Pennsylvania State University.

This work has made use of data from the European Space Agency (ESA) mission {\it Gaia} (\url{https://www.cosmos.esa.int/gaia}), processed by the {\it Gaia} Data Processing and Analysis Consortium (DPAC, \url{https://www.cosmos.esa.int/web/gaia/dpac/consortium}). Funding for the DPAC has been provided by national institutions, in particular the institutions participating in the {\it Gaia} Multilateral Agreement.

This research made use of \textsf{exoplanet} \citep{exoplanet:joss, exoplanet:zenodo} and its
dependencies \citep{celerite1, celerite2, exoplanet:agol20, exoplanet:arviz, exoplanet:astropy13, exoplanet:astropy18, exoplanet:kipping13, exoplanet:luger18, exoplanet:pymc3, exoplanet:theano}.


This research made use of Lightkurve, a Python package for Kepler and TESS data analysis (Lightkurve Collaboration, 2018).


This research has made use of the SIMBAD database,
operated at CDS, Strasbourg, France

This research has made use of the NASA Exoplanet Archive, which is operated by the California Institute of Technology, under contract with the National Aeronautics and Space Administration under the Exoplanet Exploration Program.

CIC acknowledges support by NASA Headquarters under the NASA Earth and Space Science Fellowship Program through grant 80NSSC18K1114.

\facilities{\gaia{}, HET (HPF), TESS, RBO, APO (ARCTIC), WIYN (NESSI), Shane (ShARCS), Exoplanet Archive}
\software{
\texttt{ArviZ} \citep{kumar19}, 
AstroImageJ \citep{collins17}, 
\texttt{astropy} \citep{astropy18},
\texttt{barycorrpy} \citep{kanodia18b}, 
\texttt{exoplanet} \citep{exoplanet:zenodo},
\texttt{Exo-Transmit} \citep{exotransmit.reference},
\texttt{ipython} \citep{ipython07},
 \texttt{lightkurve} \citep{lightkurve},
\texttt{matplotlib} \citep{Hunter07},
\texttt{numpy} \citep{harris20},
\texttt{pandas} \citep{reback2020pandas,mckinney-proc-scipy-2010},
\texttt{PandExo} \citep{pandexo.reference},
\texttt{PyMC3}\citep{exoplanet:pymc3},
\texttt{RadVel}\citep{fulton18},
\texttt{scipy} \citep{2020SciPy-NMeth},
\texttt{SERVAL} \citep{zechmeister18},
\texttt{starry} \citep{exoplanet:luger18},
\texttt{Theano} \citep{exoplanet:theano},
\texttt{TransitLeastSquares} \citep{hippke19}.
}

\bibliography{bibliography}

\appendix
\twocolumngrid

\clearpage

\section{Periodograms of Photometry and Activity Indicators for TOI-2136}

Due to the possibility of activity interfering with our RV measurements of TOI-2136, we include plots of several activity indicators, as well as ZTF and ASAS-SN photometry. While these data are mostly uninformative, we include them here for completeness.

\begin{figure}[p]
\centering
\includegraphics[width=0.5\textwidth]{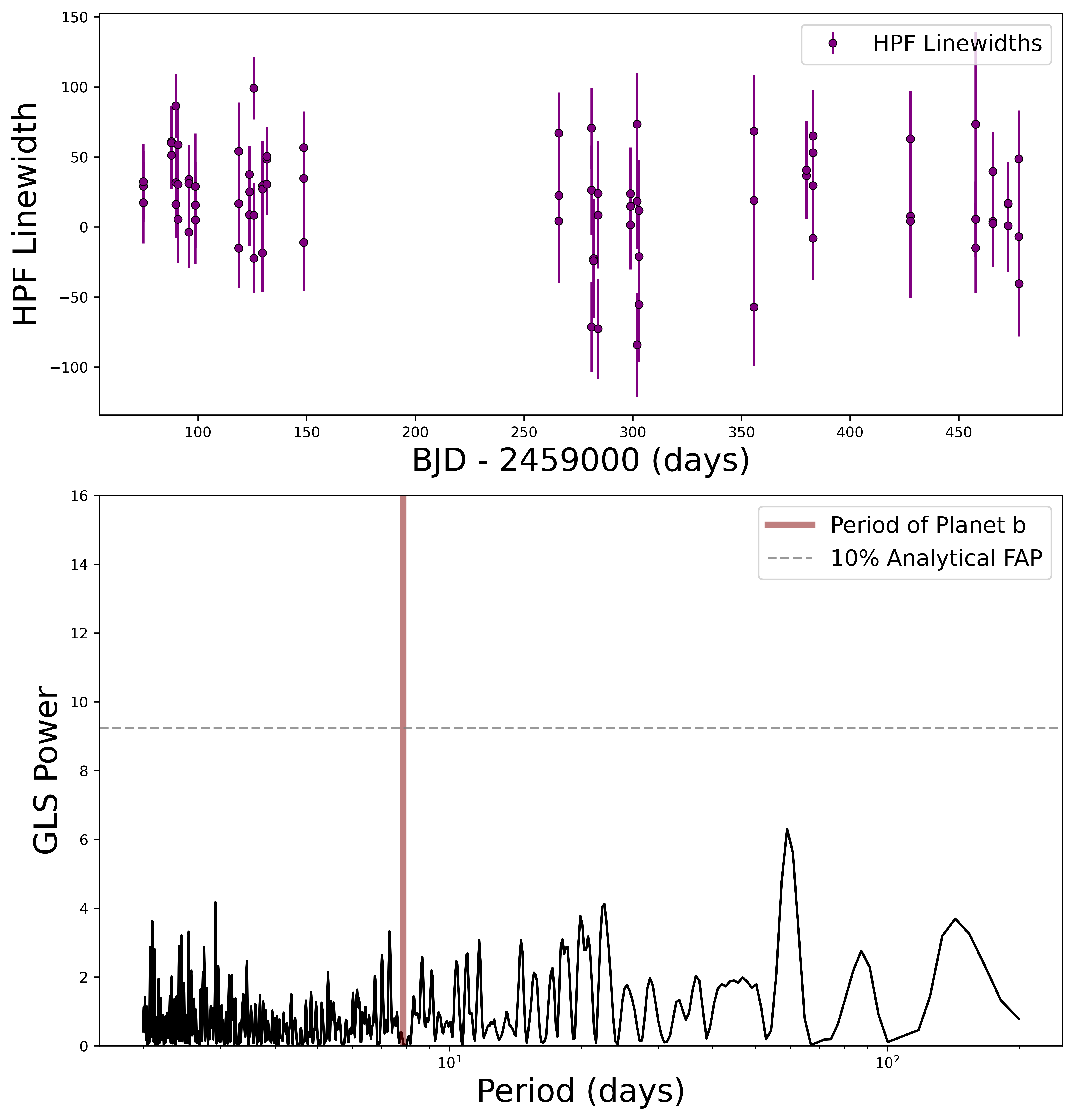}
\caption{Top: Scatterplot of HPF linewidths for TOI-2136. Bottom: GLS periodogram of TOI-2136 linewidths, with the period of planet b highlighted in red. Periodicities in linewidths can represent stellar variability, though we detect no significant periodicities in the HPF linewidths.} \label{fig:linewidth}
\end{figure}

\begin{figure}[p] 
\centering
\includegraphics[width=0.5\textwidth]{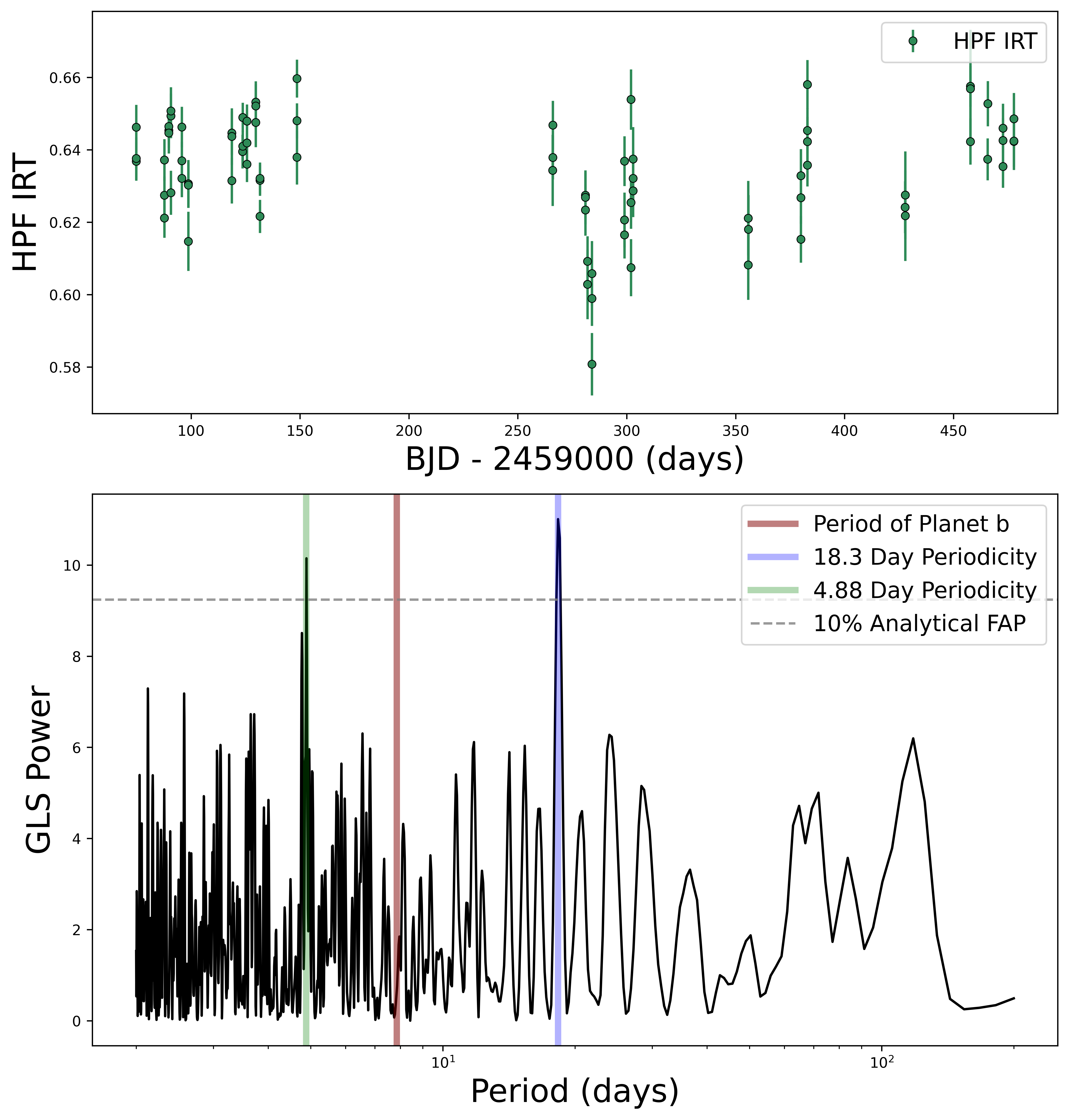}
\caption{Top: Scatterplot of the measured flux of the Calcium Infrared Triplet (Ca IRT) of TOI-2136, taken with HPF. Bottom: GLS periodogram of the data. We detect marginally strong periodicities at 4.88 days and 18.3 days, though neither seems related to the planet period, or the rotation period of the system.} \label{fig:irt}
\end{figure}

\begin{figure}[p] 
\centering
\includegraphics[width=0.5\textwidth]{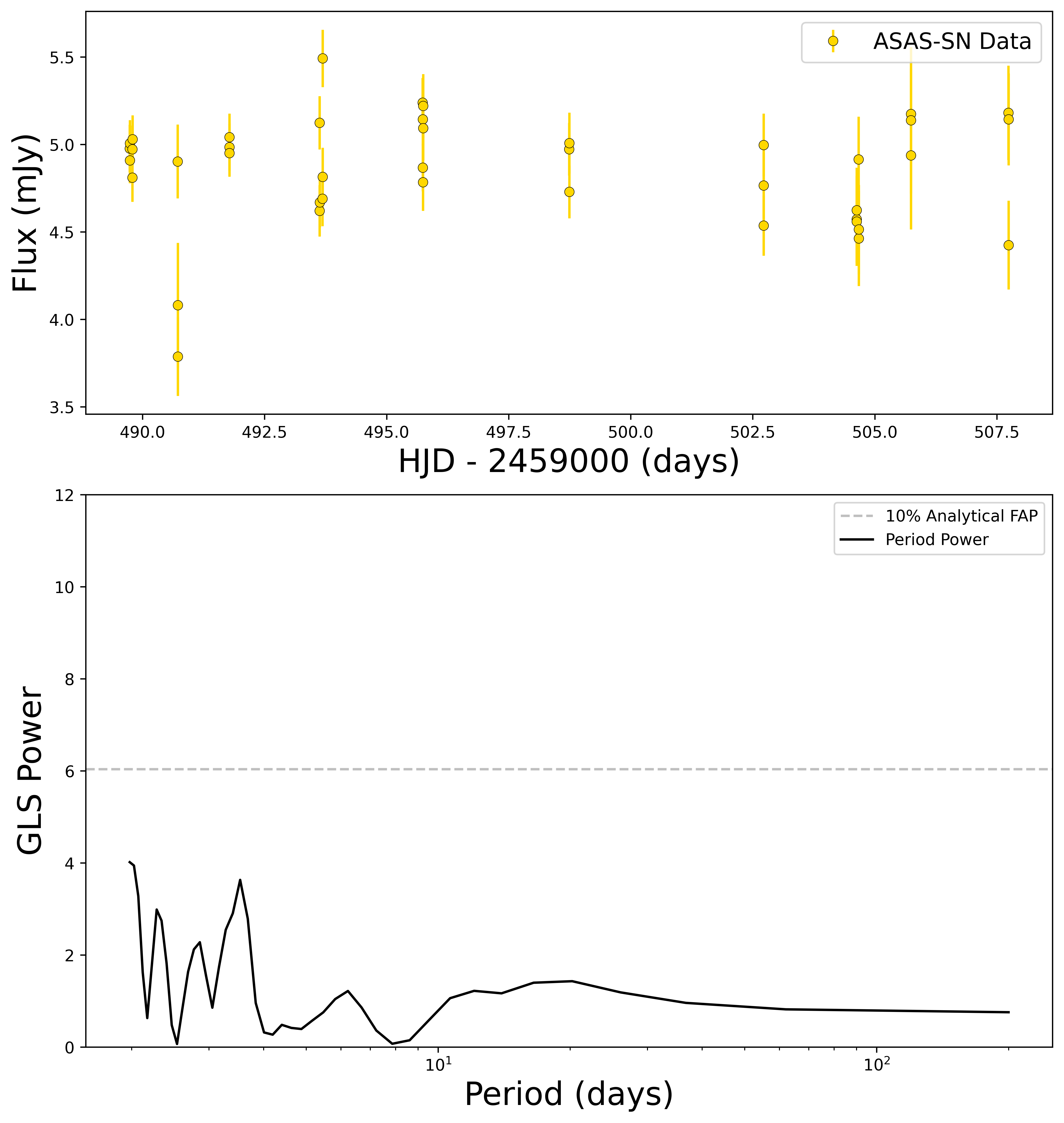}
\caption{Top: Flux data of TOI-2136 taken using ASAS-SN \citep{kochanek17}. Bottom: GLS periodogram of the data. We detect no significant periodicities.} \label{fig:asassn}
\end{figure}

\begin{figure}[p]
\centering
\includegraphics[width=0.5\textwidth]{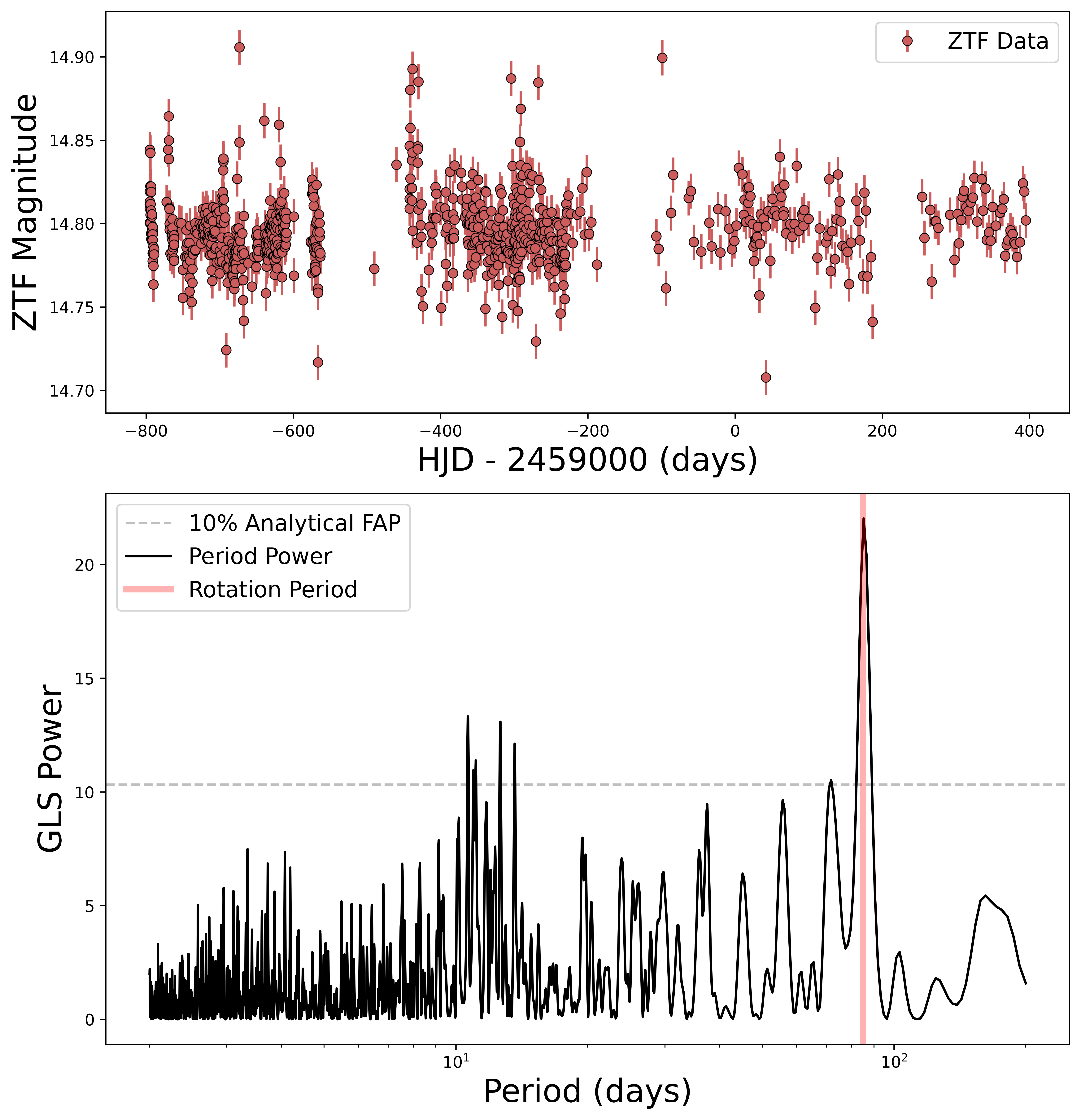}
\caption{Top: Photometric data taken using ZTF \citep{masci19}, after it has been sigma clipped for outliers. Bottom: GLS periodogram of the data. We detect a signifcant period at $\sim$ 85 days which is probably associated with the rotation of the system.} \label{fig:ztf}
\end{figure}

\end{document}